Graphical Abstract

# Direct Observation of Competing Novel Charge Density Wave Superstructures at Room Temperature in Ultrathin 1T-$VS_2$ Flakes

Samanta Pal,Kaustuv Chatterjee,Jyotirmoy Sau,Biswarup Satpati,Manoranjan Kumar,Arup Kumar Raychaudhuri,Prabir Pal

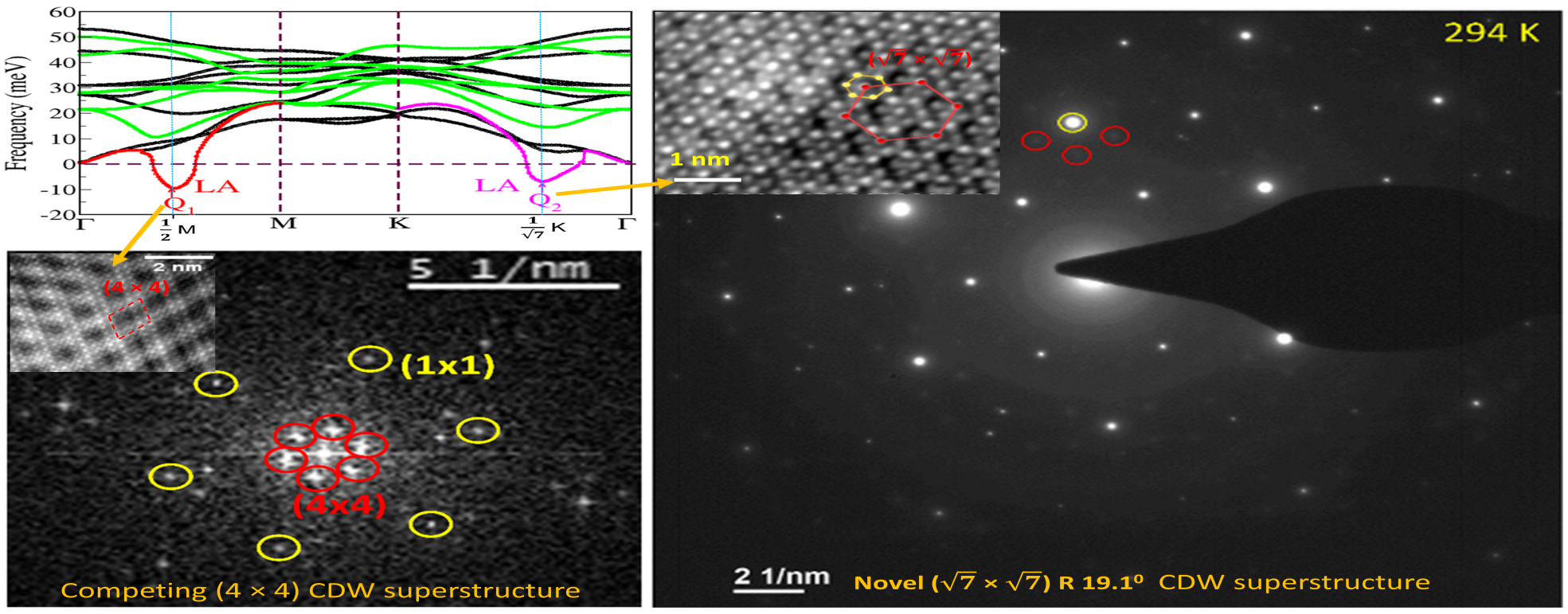

## Highlights

### Direct Observation of Competing Novel Charge Density Wave Superstructures at Room Temperature in Ultrathin 1T-$VS_2$ Flakes

Samanta Pal,Kaustuv Chatterjee,Jyotirmoy Sau,Biswarup Satpati,Manoranjan Kumar,Arup Kumar Raychaudhuri,Prabir Pal

- Experimental observation of novel CDW superstructure $(\sqrt{7} \times \sqrt{7})R$ 19.1° in ML 1T-$VS_2$ at RT
- Up to first-nearest-neighbor V-atoms cluster around a central V-atom to form the novel CDW order
- A Competition exists between the $(4 \times 4)$ and $(\sqrt{7} \times \sqrt{7})R$ 19.1° CDW superstructures in ML 1T-$VS_2$
- Monolayer $(\sqrt{7} \times \sqrt{7})R$ 19.1° CDW order can be trapped within a small-angle moiré superlattice
- A dimensional crossover to $(2 \times 2)$ CDW order is observed in multilayers of 1T-$VS_2$

# Direct Observation of Competing Novel Charge Density Wave Superstructures at Room Temperature in Ultrathin 1T-$VS_2$ Flakes

Samanta Pal[a], Kaustuv Chatterjee[a,b], Jyotirmoy Sau[c], Biswarup Satpati[d], Manoranjan Kumar[e], Arup Kumar Raychaudhuri[f,*] and Prabir Pal[a,b,*]

[a] CSIR Central Glass and Ceramic Research Institute, 196 Raja S.C. Mullick Road, Kolkata, 700032, West Bengal, India
[b] Academy of Scientific and Innovative Research (AcSIR), Sector 19, Kamla Nehru Nagar, Ghaziabad, 201002, Uttar Pradesh, India
[c] Uppsala University, Department of Physics and Astronomy, Uppsala, Box-516, SE-75120, Uppsala län, Sweden
[d] Saha Institute of Nuclear Physics, Surface Physics and Material Science Division, 1/AF, Bidhannagar, Kolkata, 700064, West Bengal, India
[e] S.N. Bose National Centre for Basic Sciences, Department of Condensed Matter Physics and Materials Science, JD Block, Sector III, Salt Lake, Kolkata, 700106, West Bengal, India
[f] UGC-DAE Consortium for Scientific Research, Kolkata Centre, Sector III, Bidhan Nagar, Kolkata, 700106, West Bengal, India

ABSTRACT

Here, we report the first experimental observation of the missing commensurate ($\sqrt{7} \times \sqrt{7}$) R19.1° charge density wave (CDW) superstructure in hexagonal 2D transition metal dichalcogenide (TMDC) prototype $VS_2$ using high-resolution transmission electron microscopy (HRTEM) and electron diffraction (ED). The clustering of the first nearest-neighbour V-atoms around a central V-atom in the hexagonal 1T-$VS_2$ lattice creates the CDW superstructure. At room temperature, the novel commensurate ($\sqrt{7} \times \sqrt{7}$) R19.1° CDW order competes with another commensurate ($4 \times 4$) CDW superstructure, attributed to distinct lattice distortions in monolayer (ML) 1T-$VS_2$. Our findings suggest that small-angle Moiré superlattices between twisted hetero-bilayers of ($1 \times 1$) and ($\sqrt{7} \times \sqrt{7}$) R 19.1° 1T-$VS_2$ do not destroy the ML CDW order but trap it within the Moiré superlattice. In multilayer $VS_2$, the in-plane competing 2D CDWs disappear due to metal ion intercalation, which leads to charge transfer and a transition to a ($2 \times 2$) type-III CDW order. Our density functional theory (DFT) calculation suggests the observed ($\sqrt{7} \times \sqrt{7}$) R 19.1° 2D CDW superstructure has lower energy ($-12.57$ meV/f.u.) compared to that of the pristine undistorted ($1 \times 1$) 1T-$VS_2$, and the softening of the low-energy acoustic phonon modes drives the CDW instability.

## 1. Introduction

The spontaneous formation of charge-density-wave (CDW) superstructures in monolayers (MLs) of a two-dimensional (2D) crystal lattice is fundamental in understanding its complex quantum states. The CDW phases of a material often compete with superconductivity [1, 2, 3, 4, 5], magnetism [6, 7], and other quantum phases [8, 3, 9, 10], and are very important in deciding the material's ground state. The origins and types of CDWs in quasi-2D materials remain debated for decades [11, 12]. Initially, it was thought that the origin of CDW in quasi-2D materials was the same electronic instability arising from Fermi surface nesting (FSN), as in 1D linear-chain compounds (Type-I CDW) [13, 14, 15, 11, 12]. But later, it was found that in quasi-2D materials, the dominant contribution comes from the structural distortion called periodic lattice distortion (PLD), which arises from momentum ($\mathbf{q}$)-dependent electron-phonon coupling (EPC), and the electronic FSN is a secondary cause of the transition (Type-II CDW) [16, 17, 18, 11, 12, 19, 20, 21, 22, 23]. The origin of CDWs in semimetals (e.g., $TiSe_2$, $HfTe_2$, etc.) [24, 25, 26] or charge-ordered (CO) materials (e.g., high-temperature superconducting cuprates) [27, 4, 28, 29, 5] cannot be fully explained only by EPC or FSN but also by other mechanisms (e.g., excitonic-condensation, electron-hole coupling, etc.) (type-III CDW).

*Corresponding author
arupraychaudhuri4217@googlemail.com (A.K. Raychaudhuri); prabir.phy@gmail.com (P. Pal)
ORCID(s): 0000-0002-8101-3297 (S. Pal); 0009-0003-4039-9327 (K. Chatterjee); 0000-0002-7624-8155 (M. Kumar); 0000-0003-1229-9532 (A.K. Raychaudhuri); 0000-0002-7356-2587 (P. Pal)

An important challenge in observing these CDW superstructures in MLs of 2D TMDC materials is the difficulty of synthesizing high-quality MLs free from defects and substrate-induced electron and hole doping, which pin and destroy the CDW phases [30, 31, 32]. One of the important factors affecting the CDW orders in multilayers of 2D transition-metal dichalcogenide (TMDC) materials is the self-intercalation of metal cations between the 2D basal planes and the resulting charge transfer. The charge transfer and resulting distortion destroy the existing in-plane 2D CDW [33, 34, 35, 36, 37, 38], but the intercalated atoms between the layers order themselves electrostatically in the octahedral or tetrahedral void positions to form interplanar charge-ordered(CO) type-III CDWs[33, 34].

Among 2D TMDCs, $TaS_2$ [39, 40, 41], $NbSe_2$ [1, 2, 42], $TiSe_2$ [43], $TiTe_2$ [44], and $VSe_2$ [45, 46] have been extensively investigated to study the formation of various types of competing CDW orders and their origins [1, 47]. Although $VS_2$ belongs to the same family of 2D TMDCs, it has been investigated only sparsely [7, 48, 49, 50, 51] due to its synthesis challenges as $VS_2$ is very sensitive to temperature and environmental oxygen species and hence may have in different non-stoichiometric phases ($V_{1+x}S_2$, $0 \leq x \leq 1$) when synthesized by high temperature chemical vapor transport or deposition techniques. The best route to get the phase pure $VS_2$ is to grow it in low-temperature ($\leq$ 300° C), inert atmospheric synthesis conditions to get rid of oxidation and sulfur vacancy or self-intercalation. CDW orders in most TMDCs generally occur at low temperatures; however, in 1T-$VS_2$, the prominent CDW superstructures form close to room temperature (300 K) [49, 52]. Observation of CDW superstructures in 1T-$VS_2$ at room temperature opens the possibility of utilizing its properties for various applications in nanoelectronics and quantum devices [53, 54, 55, 56, 57, 58]. The present investigation was carried out on $VS_2$ flakes with thicknesses ranging from ML to a few layers, obtained by liquid-phase exfoliation and subsequent stamp-transfer process (LPESTP). The motivation for conducting the experiment using the LPESTP process is to develop an alternative pathway for synthesizing samples with varying ML thicknesses while retaining these CDW and quantum properties. In this top-down LPESTP process, we do not need a substrate for ultrathin $VS_2$ growth, thereby bypassing the substrate-induced doping, strain, and Moiré patterns that can pin or destroy the CDW.

Our experimental results reveal that only 1T-$VS_2$ shows CDW formation, and the 2H-phase does not show any CDW order. The ML 1T-$VS_2$ shows a novel commensurate $(\sqrt{7} \times \sqrt{7})$ R 19.1° and a commensurate $(4 \times 4)$ CDW superstructures competing at room temperature (295 K). The $(\sqrt{7} \times \sqrt{7})$ R 19.1° CDW superstructure has not been previously found in any pristine TMDC material, including ML 1T-$VS_2$. In twisted hetero-bilayers of pristine $(1 \times 1)$ and CDW $(\sqrt{7} \times \sqrt{7})$ R 19.1° ML superstructure, a novel Moiré superlattice is formed at a twist angle 1.7° which holds the CDW periodicity of $\sqrt{7}a$. It is a special case of a small-angle Moiré trapped CDW, which was predicted earlier [59, 60, 61]. In multilayer $VS_2$, we do not observe the competing 2D CDWs of the monolayer, but instead observe a $(2 \times 2)$ type-III CDW order. The experimental investigations have been supplemented by DFT calculations to determine the origin of the novel CDW superstructures in the monolayer 1T-$VS_2$. The novel $(\sqrt{7} \times \sqrt{7})$ R 19.1° CDW phase in our freestanding flakes has been achieved through exfoliation and strain stabilization. This approach circumvents substrate-induced charge transfer and epitaxial constraints, providing a unique framework to explore new CDW states that are challenging to observe in strongly pinned epitaxial films. Our findings establish ML 1T-$VS_2$ synthesized via top-down LPESTP as a unique platform for constructing and investigating new CDW orders.

## 2. Materials and methods

### 2.1. Liquid Phase Exfoliation and stamp transfer process (LPESTP) of $VS_2$ flakes

The precursor bulk $VS_2$ samples have been synthesized via the hydrothermal process, as reported elsewhere [52]. To obtain ultrathin $VS_2$ flakes with single crystalline domains from this oriented bulk $VS_2$ sample, we used the combined liquid phase exfoliation technique [62, 63] and the subsequent stamp transfer process (LPESTP) [64] as shown schematically in Figure 1. Here, 1 mg of oriented bulk $VS_2$ powder was mixed with 20 ml of liquid formamide (FA) in a capped glass cuvette and ultrasonicated for 8-9 hours in a bath sonicator in an ice-water bath to obtain very thin exfoliated $VS_2$ flakes. After ultrasonication of $VS_2$ in FA solution, we performed vacuum suction filtration of the process on a PTFE membrane with a 0.2 $\mu$m pore size (Whatman, Cytiva, Cat. No. 7582-004, Japan). A mixed cellulose membrane with the same pore size (0.2 $\mu$m pore, Whatman, Cytiva, Product No. 10401712, Germany) is placed on top of the PTFE membrane and is roll-pressed by hand with a glass rod. The mixed cellulose membrane was then separated from the PTFE membrane to obtain the transferred ultrathin $VS_2$ flakes attached to the mixed cellulose membrane. These ultrathin $VS_2$ flakes were then transferred to Cu grids and $SiO_2$ (300 nm) / Si (100) substrates by placing small pieces of them. Cellulose was removed by dipping it in acetone overnight and then drying it at RT in a clean room, yielding good quality, substrate-induced doping- and strain-free ultrathin $VS_2$ nanoflakes.

### 2.2. Characterizations and Measurements

We have characterized the synthesized bulk precursor $VS_2$ as well as ultrathin and multilayer $VS_2$ flakes obtained from the LPESTP method. Transmission electron microscopy (TEM) at room temperature was performed using a Tecnai $G^2$ 30ST microscope operated at 300 kV (FEI, Netherlands/currently Thermo Fisher Scientific, USA) and equipped with facilities for selected area electron diffraction (SAED) and energy dispersive spectroscopy (EDS). The Atomic Force Microscopy (AFM) topographies of the LPESTP transferred $VS_2$ flakes are measured in tapping mode with MFP-3D Origin+ Asylum Research AFM by Oxford Instruments. X-ray photoelectron spectroscopy (XPS) measurements were made using a PHI 5000 Versa Probe-II under UHV at a base pressure of 2.3 x $10^{-10}$ mbar. The samples were grounded to the platen with copper tabs, ensuring minimal surface charge accumulation during X-ray irradiation. Additionally, a low-energy electron flood gun was used alongside an $Ar^+$ ion gun operated at a low accelerating voltage for effective surface charge neutralization. A focused X-ray ($\approx 100\mu$m) source of Al $K_\alpha$ (1486.7 eV) was used for X-ray photoemission studies, while the valence band was investigated using a He-I VUV source (21.2 eV) with an approximate beam size of ($\approx 500\mu$m). The survey spectra were recorded with the analyzer pass energy set at 187.85 eV and a step size of 0.2 eV; for the core levels, the pass energy was adjusted to 11.57 eV with a step size of 0.05 eV, achieving a maximum resolution of 300 meV. The UPS spectra were acquired at a pass energy of 2.95 eV and a step size of 10 meV, reaching a maximum resolution of 50 meV. The carbon peak [C-C ($sp^3$)] was calibrated at 284.7 eV for charge correction before analysis. The valence band spectra were calibrated against the Fermi edge of an in situ sputter-cleaned Au foil. Temperature-dependent resistivity measurements were performed by a four-linear probe method on pressed pellets made with microflakes of $VS_2$ using a low-temperature closed-cycle cryostat system down to 10 K. The data were taken during the heating cycle with a heating rate of 2 K/min as programmed by the temperature controller. The simulated Moiré superlattice pattern is constructed using the Twister software package by introducing a twist angle of 1.7° between $(1\times1)$ and $(\sqrt{7}\times\sqrt{7})$ R 19.1° $VS_2$ monolayers with AA stacking. Computational details of the software can be found elsewhere [65].

### 2.3. Computational Details

For the electronic ground-state structure calculations, the *projector augmented-wave* (PAW) method [66] efficiently combines the computational simplicity of pseudopotential-based approaches [67] with the accuracy of all-electron methods by reconstructing the full valence wavefunctions from the pseudo-wavefunctions. A plane-wave energy cutoff of 600 eV was used to ensure accurate convergence of the total energy. The *self-consistent field* (SCF) calculations were performed with a stringent energy convergence criterion of $10^{-8}$ eV between successive electronic steps. The Brillouin zone (BZ) was sampled using a Monkhorst–Pack $k$-point grid for the primitive unit cell to achieve precise electronic structure results, while maintaining an equivalent $k$-point density for all supercell calculations. We have used calculations for undistorted 1T-$VS_2$ structure, 2H, $(\sqrt{7}\times\sqrt{7})\,R\,19.1°$, $(7\times\sqrt{3}\,R\,30°)$ CDW orders with 5% lattice distortion, consistent with our TEM analysis.

For the phonon dispersion and vibrational property calculations, the finite-displacement method [68, 69] was employed within the supercell approach. We employed the projector augmented wave (PAW) method [66] based on a plane-wave basis set within the framework of *density functional theory* (DFT). The *exchange–correlation energy* was treated using the *generalized gradient approximation* (GGA) in the *Perdew–Burke–Ernzerhof* (PBE) parameterization [67], as implemented in the VASP code [70]. A $(4\times4\times1)$ supercell of the conventional unit cell for pristine $VS_2$, which consists of 48 atoms, was constructed to accurately capture the long-range interatomic force constants. Each atom in the supercell was displaced by 0.01 Å from its equilibrium position to compute the interatomic force constants from the DFT-calculated forces. On the other hand, for the $(\sqrt{7}\times\sqrt{7})\,R\,19.1°$ configuration, we used a $(2\times2\times1)$ supercell consisting of 84 atoms, while keeping all other parameters identical to those used for the pristine structure. These force constants were subsequently used to determine the phonon frequencies and dispersion relations, providing insights into the dynamical stability and lattice dynamics of the system.

## 3. Results and discussion

We can transfer ultrathin atomic layers of $VS_2$ flakes onto any desired substrate using the LPESTP method, as shown in the schematic of Figure 1(a). Ultrathin layers, as observed in atomic force microscopy (AFM), have a thickness (t) ranging from ML to 10-11 layers ($0.6 \leq t \leq 7\ nm$), and lateral sizes (L) ($0.2 \leq L \leq 1\ \mu m$), making them suitable for direct observation of CDWs in high-resolution transmission electron microscopy (HRTEM) and electron diffraction (ED) experiments which are considered two of the best techniques for directly observing CDWs with the highest resolution [40, 41]. The height measurements by AFM (Figure S3, Support Information) indicate that the thinnest $VS_2$ flakes obtained have a typical thickness of 0.65 nm, corresponding to the ML $VS_2$ [71, 72, 73].

### 3.1. Polymorphic phase transitions in ultrathin $VS_2$ flakes

An interesting observation is the coexistence of the 1T, 2H, and 1T' polymorphs in monolayer $VS_2$ as shown in Figures 1 (b) and (c). TEM is one of the best techniques for identifying these polymorphic phases [74, 75]. The 1T, 2H, and 1T' polymorphs of $VS_2$ are marked by red, yellow, and green lines or squares, respectively. Inset of (b) shows the fast Fourier transform (FFT) of the above real-space HRTEM image to get the reciprocal-space spots and symmetry. The clear, intense hexagonal spots indicate the single-crystalline nature and hexagonal lattice phase of the $VS_2$ flake. Figure 1 (c) shows the HRTEM image of a portion of the flake showing different regions of polymorphic phase transitions as marked. In Figures 1(d) and 1(e), 2H and 1T' polymorphs are respectively identified from the atomically resolved HRTEM images. Atomic-structure models of the polymorphic phases, along with HRTEM images, confirm their coexistence in the ML $VS_2$ flakes. The transition between the 2H and 1T polymorphic phases in $VS_2$ is commonly attributed to the lattice strain and charge transfers [75, 76]. The lattice strain induced by charge transfer destabilizes the 2H phase (trigonal-prismatic) and stabilizes the 1T phase (octahedral). The transition from the 2H to the 1T phase in $VS_2$ can be visualized as a gliding or lateral shift of one of the S planes within the S-V-S trilayers. This shift causes the central vanadium (V) atom to become symmetrically surrounded by six sulfur atoms, forming an octahedrally coordinated environment, characteristic of the 1T phase. Structurally, this 2H-to-1T transformation is achieved by gliding one of the sulfur planes along the $\langle 2100 \rangle$ crystallographic direction by a magnitude of $a/\sqrt{3}$, where a is the in-plane lattice constant [74, 75]. A further distortion of the 1T polymorph gives rise to the 1T' phase, which features a distorted octahedral coordination or zigzag chains of vanadium (V) atoms within the structure as a result of Jahn-Teller distortion, which lifts the degeneracy and lowers the symmetry and energy of such metal complexes having unpaired d-orbitals [35].

### 3.2. ML $VS_2$ and observation of a novel ($\sqrt{7} \times \sqrt{7}$) R19.1° CDW superstructure

Figure 2(a) shows a real-space atomic-resolution HRTEM image of an ML 1T- $VS_2$ flake. In the top-right inset, clear views of the pristine hexagonal Bragg lattices and the modulated CDW superlattices in real space are marked by yellow and red dots, respectively. In the bottom-right inset, the fast Fourier transform (FFT) of the HRTEM image is shown, providing reciprocal-space information of the Bragg and CDW superlattice orders. The Bragg and CDW spots are marked by yellow and red circles, respectively. The unit cell of the CDW superstructure makes an angle of 19.1° with the Bragg lattice as seen in the FFT.

The atomically resolved HRTEM image and its FFT spots confirm the novel commensurate ($\sqrt{7} \times \sqrt{7}$) R 19.1° CDW superstructure in ML 1T-$VS_2$. We observed this missing novel CDW superstructure for the first time in ML 1T-$VS_2$ and also in the whole family of pristine TMDC materials, although it was predicted to be present when first-nearest-neighbor metal atoms (n=1, where n is the no. of shells or nearest-neighbor no. surrounding a central atom) cluster around a central metal atom [59, 77, 78]. Our novel CDW formation involves a periodic lattice distortion, and as a result, we get clear atomic "grouping" or "clustering" of the first six nearest-neighbor atoms to a central atom, i.e., a total of seven (6n+1, where n= 1) atom cluster, similar to the formation of a star of David (SoD) in $TaS_2$ [40]. However, in the case of $TaS_2$, total twelve Ta-atoms cluster around a central Ta-atom (6n+1, where n= 2 for the first and second nearest neighbors) to form the ($\sqrt{13} \times \sqrt{13}$) R 13.9° SoD and in the case of 1T-$TaTe_2$ eighteen Ta-atoms cluster around a central Ta-atom (6n+1, where n= 3) to obtain the ($\sqrt{19} \times \sqrt{19}$) R 23.4° CDW superstructure (an extended SoD) [79]. Here, in the case of $VS_2$, the first six (first nearest, n= 1) V-atoms move toward a central V-atom to form a seven-atom cluster and are surrounded by similar six-hexagonal clusters of seven 'V-atoms groupings' to make a tight hexagonal CDW superstructure, as marked by a red hexagon, as shown in Figure 2(a). The yellow hexagon marked forms the Bragg lattice. Two unequal V-V bond lengths resulting from the periodic lattice distortion in the CDW superstructure have been found from the line profile data of the HRTEM images as depicted in Figure S4 and Table S1 (Support Information). The short bond-length ($V\text{-}V_{short} = 3.01 \pm 0.02$ Å) is between the central V-atom and the first nearest V-atoms, and the long bond-length ($V\text{-}V_{long} = 3.34 \pm 0.02$ Å) is between the two nearest collinear intercluster V-atoms. A comparison of CDW orders in ML TMDCs is presented in Table S2 (Supporting Information).

The calculated in-plane lattice parameter of the Bragg lattice unit cell is $a_{lattice} = 3.18 \pm 0.02$ Å, while the lattice parameter of the unit supercell of the CDW superstructure is $a_{CDW} = \sqrt{7} a_{Bragg} = 8.41 \pm 0.02$ Å. The ML 1T-$VS_2$ has one V atom per unit cell, while the $(\sqrt{7} \times \sqrt{7})$ R 19.1° CDW superstructure has seven (6n+1, where n=1) V atoms per unit supercell.

In Figure 2 (b), the schematics of the unit cells and first Brillouin zones for the (1 × 1) pristine and $(\sqrt{7} \times \sqrt{7})$R 19.1° CDW order, as marked by yellow and red lines, are shown, respectively, in reciprocal space along with the CDW modulation positions (red points). To reconstruct the $(\sqrt{7} \times \sqrt{7})R19.1°$ CDW superstructure from a hexagonal parent lattice, a convenient choice of real-space supercell basis vectors is $\mathbf{A}_1 = 2\mathbf{a}_1 + \mathbf{a}_2$ and $\mathbf{A}_2 = -\mathbf{a}_1 + 3\mathbf{a}_2$, where $\mathbf{a}_1$ and $\mathbf{a}_2$ denote the real-space basis vectors of the pristine (1 × 1) 1T-$VS_2$ lattice and $|\mathbf{a}_1| = |\mathbf{a}_2| = |\mathbf{a}|$. The superstructure basis vectors have a magnitude of $\sqrt{7}a$ and create a rotation of $\theta = \tan^{-1}(\sqrt{3}/5)$, approximately 19.1°, relative to the original lattice. The corresponding reciprocal superlattice vectors are given by $\mathbf{B}_1 = \frac{3}{7}\,\mathbf{b}_1 + \frac{1}{7}\,\mathbf{b}_2$ and $\mathbf{B}_2 = -\frac{1}{7}\,\mathbf{b}_1 + \frac{2}{7}\,\mathbf{b}_2$, along with their symmetry-related counterparts. The reciprocal-space basis vectors of the unaltered (1 × 1) 1T-$VS_2$ are represented here by $\mathbf{b}_1$ and $\mathbf{b}_2$, where $|\mathbf{b}_1| = |\mathbf{b}_2| = |\mathbf{b}|$. The magnitude of $\mathbf{B}_i$ is expressed as $|\mathbf{B}_i| = \frac{1}{\sqrt{7}}|\mathbf{b}| = \frac{4\pi}{\sqrt{21}a} \approx 0.378\ |\mathbf{b}|$. Furthermore, the angle between $\mathbf{B}_1$ and $\mathbf{B}_2$ is 120°. The $(\sqrt{7} \times \sqrt{7})$ reconstruction is defined by a two-dimensional matrix of the reciprocal-space vectors as mentioned above. Each rotated supercell domain produces six modulation peaks $(\pm\mathbf{B}_1, \pm\mathbf{B}_2, \pm\mathbf{B}_3)$, situated at a radius of $\frac{1}{\sqrt{7}}\ |\mathbf{b}|$ from the Γ point and rotated by 19.1° from the high symmetry $\Gamma - M$ axes of the pristine Bragg lattice 1T-$VS_2$.

Figure 2(c) shows the schematic illustration of the $(\sqrt{7} \times \sqrt{7})$ R 19.1° CDW (marked by red) and corresponding undistorted (1 × 1) Bragg lattice (marked by yellow) unit cells showing the relative angle and surface re-arrangements. Figures 2 (d), 2(e), and 2(f) show the temperature-dependent (294 K, 218 K, and 123 K, respectively) selective area electron diffraction patterns taken from such an ultrathin $VS_2$ flake. We can clearly see six hexagonally arranged CDW superstructure spots (marked three with red circles) around each bright Bragg spot (marked in yellow) in the SAED pattern taken at 294 K. These superspots vanish at 218 K and 123 K and reappear at 294 K during reheating (Figure S12, Support Information), indicating a CDW transition near room temperature ($T \approx 295\ K$) and its melting at low temperatures ($T \leq 200\ K$). The hexagonally arranged superspots, which vanish with decreasing temperature, rule out the possibility that they arise from overlapping multilayers or from any other symmetry-lowering phenomena.

### 3.3. (4×4) CDW Orders and DFT Calculations for the Competing CDW Orders in ML $VS_2$

We have observed a competing commensurate (4 × 4) CDW order at room temperature in ML $VS_2$ with ≈ 2.51% lattice distortion (with distorted short and long bond-length, $V\text{-}V_{short} = 3.10$ Å and $V\text{-}V_{long} = 3.24$ Å; and the undistorted (1×1) V-V bond-length = 3.18 Å) as shown in Figures 3(a) -(c). Figure 3(a) shows the HRTEM image of a flake consisting of both the pristine (1 × 1) 1T-$VS_2$ and (4 × 4) CDW superstructure marked by regions I and II, respectively. Figures 3(b) and (c) show the HRTEM image and FFT of region II, respectively. In Figure 3(b), a unit cell of the (4×4) superstructure in real space is marked by a red rhombus, and in Figure 3(c), the Bragg and superspots are marked in reciprocal space by yellow and red circles, respectively. Similarly, Figures 3(d) and 3(e) show the HRTEM image and FFT of region I. In 3(d), a unit cell of the pristine (1 × 1) Bragg lattice is marked in real space by a yellow rhombus, and in 3(e), the pristine Bragg spots in reciprocal space are marked by yellow circles. We have seen that these two competing CDW orders are stabilized with different lattice distortions. The (4 × 4) CDW superstructure wins for low atomic distortions (≤ 4%); however, at higher atomic distortions (≥ 4%), the $(\sqrt{7} \times \sqrt{7})$ R 19.1° CDW superstructure wins. Previously, Lee et al. observed a (4×4) CDW superstructure in few-layer 1T-$VS_2$ [51]. Monolayer $VTe_2$ and $NbTe_2$ also exhibit a (4 × 4) CDW superstructure [80, 81]. DFT calculations indicate a (4 × 4) ground state for monolayer 1T-$VS_2$ and related materials, driven by Fermi surface nesting at q = 1/2 $|\Gamma - M|$ [7, 82]. We found that liquid-phase exfoliation and stamp transfer (LPESTP) produces ultrathin flakes without substrate fixation. The nesting-induced (4 × 4) superstructure is facilitated by areas with lower atomic distortion (lattice strain ≤ 4%), while adjacent regions with higher local strain (≥ 4%) exhibit the more complex $(\sqrt{7} \times \sqrt{7})$ R 19.1° CDW superstructure.

We investigated the relative stability of various $VS_2$ monolayer structures with respect to the pristine (1×1) 1T-$VS_2$ reference structure. The relative energy per formula unit ($\Delta E$) was evaluated as

$$\Delta E = \frac{E_{\text{supercell}}}{N_{\text{f.u.}}} - E_{\text{ref}}, \tag{1}$$

where $E_{\text{supercell}}$ is the total energy of the supercell, $N_{\text{f.u.}}$ is the number of formula units within the supercell, and $E_{\text{ref}}$ is the energy per formula unit of the $(1 \times 1)$ 1T-$VS_2$ monolayer.

**Table 1**
Total energy difference $\Delta E$ (meV/f.u) for $(1 \times 1)$ ML 2H-$VS_2$, $(\sqrt{7} \times \sqrt{7})$ R 19.1°, $(4 \times 4)$ and $(7 \times \sqrt{3}$ R 30°) CDW ML supercells with respect to the pristine $(1 \times 1)$ ML 1T $VS_2$.

| Material Phase | Monolayer Superstructures | $\Delta E$ (meV/f.u.) |
|---|---|---|
| ML 2H-$VS_2$ | $(1 \times 1)$ | $-3.00$ |
| ML 1T-$VS_2$ | $(4 \times 4)$ | $-11.16$ |
| ML 1T-$VS_2$ | $(\sqrt{7} \times \sqrt{7})\, R\, 19.1°$ | $-12.57$ |
| ML 1T-$VS_2$ | $(7 \times \sqrt{3}\, R\, 30°)$ | $-11.84$ |

Table 1 presents the relative ground-state energies of the considered $VS_2$ monolayer configurations. The monolayer 2H-$VS_2$ phase exhibits a relative energy change $\Delta E = -3$ meV/f.u., indicating a slightly higher stability than the monolayer 1T-$VS_2$ phase. The $(4 \times 4)$ ML CDW supercell (with 2.5 % lattice distortion) shows energy reduction of $-11.16$ meV/f.u., while the novel superstructure $(\sqrt{7} \times \sqrt{7})$ R 19.1° (with 5.3 % lattice distortion) and the $(7 \times \sqrt{3}$ R 30°) superstructure [7], exhibit significantly reduced energies of approximately $-12.57$ meV/f.u. and $-11.84$ meV/f.u., respectively, demonstrating enhanced stability relative to both the (1×1) 1T, 2H monolayers and (4×4) CDW supecells. This means that rotating and rearranging the lattice makes the system more stable. In other words, the way the atoms shift and rearrange helps the $VS_2$ monolayer settle into its charge-density-wave (CDW) state.

The onset of CDW order in the quasi-2D materials is related to the lattice instabilities [23, 17, 19, 21, 22]. We show that the emergence of the CDW order is driven by softening of phonon modes, as described below. To elucidate the vibrational properties, the experimentally observed crystal structures were employed to compute the dynamical matrices of the $(1 \times 1)$ and $(\sqrt{7} \times \sqrt{7})$ R 19.1° 1T-$VS_2$ monolayers. Phonon calculations were carried out within the harmonic approximation, where the dynamical behaviour of atoms is governed by the equations

$$D(\mathbf{q})\, \mathbf{e}_{\mathbf{q}j} = \omega^2_{\mathbf{q}j}\, \mathbf{e}_{\mathbf{q}j}, \tag{2}$$

with $\mathbf{q}$ and $j$ denoting the phonon wavevector and mode index, respectively. Here, $\omega_{\mathbf{q}j}$ and $\mathbf{e}_{\mathbf{q}j}$ represent the phonon frequency and polarization vector of the corresponding mode. The Hermitian nature of the dynamical matrix $D(\mathbf{q})$ ensures that all eigenvalues $\omega^2_{\mathbf{q}j}$ are real.

The phonon density of states (DOS) is defined as

$$g(\omega) = \frac{1}{3N} \sum_{\mathbf{q},j} \delta(\omega - \omega_{\mathbf{q}j}), \tag{3}$$

where $N$ is the total number of unit cells. The total DOS is normalized such that $\int g(\omega)\, d\omega = 3n$, where $n$ denotes the number of atoms per unit cell.

The *ab initio* density functional perturbation theory (DFPT) calculations for the $(\sqrt{7} \times \sqrt{7})$ R 19.1° reconstructed 1T-$VS_2$ monolayer reveal a clear structural instability that drives the system toward a charge density wave (CDW) ground state. As shown in Figure 3(f), the phonon dispersion exhibits pronounced imaginary modes across select regions of the Brillouin zone. Significant softening appears along both the Γ-K and Γ-M directions, with the dominant $Q_1$ instability located near the midpoint of $\Gamma - M$ ($q = \frac{1}{2}|\Gamma - M|$) and the $Q_2$ mode positioned approximately at $\frac{1}{\sqrt{7}}|\Gamma\text{–K}|$. The $Q_1$ and $Q_2$ modes are associated with low-energy longitudinal acoustic vibrations.

We have performed structural relaxations on the ($\sqrt{7}\times\sqrt{7}$) R 19.1° unit cells for the harmonic DFPT calculations. The ($\sqrt{7}\times\sqrt{7}$) R 19.1° unit cell was constructed by displacing the vanadium atoms from their symmetric positions by the experimentally calculated value of 5.3% of the lattice parameter, while the positions of the sulfur atoms remain almost unchanged. The associated energy gain is about 12.57 meV. We find an instability of the longitudinal acoustic branch approximately at the wavevector q = $\frac{1}{\sqrt{7}}|\Gamma$–K$|$. However, the dominant instability within the harmonic approximation is located at q=1/2 $|\Gamma - M|$ in the longitudinal–acoustic branch. The $Q_1$ anomaly located at q = 1/2 $|\Gamma - M|$, related to the instabilities in the longitudinal acoustic branch arising mainly from Fermi surface nesting, is commensurate with the superstructure (4×4) and related to small ($\leq$ 4%) atomic distortions [7, 82, 83]. The $Q_2$ anomaly presented in Figure 3(f) originates from a one-dimensional high-symmetry cut along the $\Gamma - K$ direction, positioned at 0.370 $|\Gamma - K|$ (approximately $\frac{1}{\sqrt{7}}$ $|\Gamma - K|$). The calculated $Q_2$ phonon-softening point ($\approx \frac{1}{\sqrt{7}}$ $|\Gamma - K|$) is also evident in the attenuation of the longitudinal acoustic (LA) mode of monolayer 1T-$VS_2$, as observed in previous studies [7, 84]. However, this calculated $q$-value differs from the empirically derived $q$-value of $\frac{1}{\sqrt{7}}|b| = \frac{\sqrt{3}}{\sqrt{7}}\, cos(10.9°)\, |\Gamma - K|$, which is approximately 0.64 $|\Gamma - K|$. This finding is consistent with earlier reported values on similar two-dimensional CDW materials such as $TaS_2$ and $TaTe_2$ [79, 85]. The CDW superstructures identified in 2D $TaS_2$ and $TaTe_2$ in monolayer form are $(\sqrt{13}\times\sqrt{13})R13.9°$ and $(\sqrt{19}\times\sqrt{19})R23.4°$, respectively. DFT harmonic stabilization analysis showed phonon mode softening at $\frac{1}{\sqrt{13}}$ (approximately 0.277) in the Γ–K direction. However, the calculated $q$-value (approximately 0.277$|\Gamma - K|$) differs from the actual experimental projected values of $\frac{1}{\sqrt{13}}|\, \mathbf{b}| = 0.462\ |\Gamma - K|$ and $\frac{1}{\sqrt{19}}|\, \mathbf{b}| = 0.395\ |\Gamma - K|$ along the $\Gamma - K$ direction for the $(\sqrt{13}\times\sqrt{13})R13.9°$ in $TaS_2$ and $(\sqrt{19}\times\sqrt{19})R23.4°$ in $TaTe_2$ CDW superstructures, respectively [79, 85]. This is consistent with our DFPT computations. The exact mode-softening minima along these high-symmetry axes in the harmonically stabilized configuration seldom coincide with the $q$-vectors associated with off-axis aligned significantly restructured superstructures. However, to get the experimental $q$-value with the phonon mode softening predicted by DFPT calculations, significant lattice distortion may require higher-order anharmonic displacement terms [7]. Consequently, rather than representing the pure harmonic ground state, the commensurate $(\sqrt{7}\times\sqrt{7})R19.1°$ CDW superstructure should be considered an anharmonically stabilized metastable structural phase of pristine 1T-$VS_2$.

The pristine (1×1) 1T-$VS_2$ monolayer shows no phonon mode softening throughout the Brillouin zone, confirming its dynamical stability and the absence of CDW-related distortions. The overall reduction in phonon energies for the ($\sqrt{7}\times\sqrt{7}$) R 19.1° system reflects the redistribution of vibrational energy as the structure relaxes into the CDW phase. The phonon density of states (PDOS), shown in Figure 3(g), further supports this interpretation. In the reconstructed monolayer, negative-frequency contributions originate mainly from vanadium vibrations, indicating that the instability is driven by the V sublattice.

## 3.4. Moiré trapped CDW in twisted bilayer of 1T-$VS_2$

Moiré pattern is one of the more interesting atomic arrangements formed in multilayered 2D materials. This generally occurs when two periodic structures with similar but not identical periodicities or orientations are overlaid on each other [86]. When an additional layer is overlaid on the CDW monolayer, it typically alters or destroys the CDW pattern due to interlayer coupling and charge transfer [44, 87, 30]. It is an interesting idea to investigate what happens to our monolayer CDW pattern when a (1×1) 1T-$VS_2$ layer is overlaid on top of the ($\sqrt{7}\times\sqrt{7}$) R 19.1° CDW layer. We show below such a small-angle heterobilayer Moiré superlattice pattern.

Figure 4 (a) shows the Moiré superlattice patterns in the twisted hetero-bilayer of (1 × 1) 1T-$VS_2$ and ($\sqrt{7}\times\sqrt{7}$) R 19.1° CDW layer at a twist angle of 1.7°. Figure 4(b) is the FFT of 3(a), showing clearly the Bragg spots of the two layers, making a (19.1 + 1.7) = 20.8° angle with each other as marked by red and yellow circles. Violet circles mark both the CDW and periodic Moiré spots in reciprocal space, as the Moiré periodicity is the same as the CDW periodicity and the Moiré pattern holds the periodic order of ($\sqrt{7}\times\sqrt{7}$). It is thus clear that the ($\sqrt{7}\times\sqrt{7}$) R 19.1° CDW order holds in the small-angle Moiré bilayer. This is a case of a small-angle Moiré-trapped CDW in 2D TMDC materials, as predicted earlier [59, 60, 61]. The difference between the periodic ($\sqrt{7}\times\sqrt{7}$) R 19.1° CDW and periodic Moiré ($\sqrt{7}\times\sqrt{7}$) is that in monolayer there is clustering of 6 V-atoms around a central atom i.e. in total 7 atoms cluster together to form the unit cell of the CDW lattice (such as formation of SOD in $TaS_2$) but in bilayer Moiré there is rotation of 7 atoms with respect to the other 7 atoms i.e. a unit Moiré cell consists of 14 V-atoms. In the Moiré pattern, one can get in SAED two types of Bragg spots separated by an angle related to the twist and Moiré periodic spots, but in the CDW superstructure, one can get one type of Bragg spots surrounded by symmetric superstructure spots. Figure 4(c) is the simulated Moiré pattern between a hetero-bilayer of $(1 \times 1)$ 1T-$VS_2$ and $(\sqrt{7} \times \sqrt{7})$ R 19.1$^o$ CDW layer at a twist angle of 1.7$^o$ that closely matches our experimental Moiré patterns. This confirms that monolayer $(\sqrt{7} \times \sqrt{7})$ R 19.1$^o$ CDW does not get destroyed in the small-angle twisted bilayer and rather gets trapped in the Moiré bilayer.

## 3.5. Multilayer $VS_2$ and appearance of $(2 \times 2)$ CDW order

We have also observed a $(2 \times 2)$ CDW order in multilayer $VS_2$ flakes, as previously reported by a few groups in multilayer intercalated TMDC materials [48, 34, 33]. The origin of this type of CDW is assumed to be the electrostatic charge ordering of the self-intercalated metal cations ($M^{\delta+}$, $3 \leq \delta < 4$) in the interlayer void positions of 2D TMDC materials $M_{1+x}X_2$, where M = transition metal, S = S, Se or Te and x is a fractional number which indicates the intercalation amount. This type of charge ordering in the 2D plane of layered materials gives a charge density variation [41, 5] and refers to the type-III CDW category [11, 12].

Figure 5(a) shows the electron diffraction (ED) pattern of a multilayer $VS_2$ flake taken at room temperature for direct reciprocal space information. The CDW superstructure spots are marked in red, whereas the Bragg lattice spots are marked in yellow. A closer view of the diffraction spots near the central beam is shown in the inset of the Figure. It is evident that the distance of the superstructure spots from the central spot in reciprocal space is half of that between the Bragg and central spot and does not create an angle with the Bragg-lattice spots (affirming its commensurate nature). The 2D atomic arrangement of this charge order in multilayer $VS_2$ in real space is $(2 \times 2)$ R0$^o$. Figure 5(b) is an HRTEM image of a multilayer $VS_2$ flake showing a clear commensurate $(2 \times 2)$ superstructure order. The top-right inset shows the FFT spots corresponding to the HRTEM image, clearly showing the $(2 \times 2)$ order spots and Bragg spots, marked by red and yellow circles, respectively. The bottom-right inset shows a zoomed-in view of the region marked by the white border, where pristine Bragg lattice points (yellow) and superlattice points (red) are shown. The distance between the two nearest vanadium atoms for the Bragg lattice is half the distance for the $(2 \times 2)$ superlattice. This has been verified by measuring the corresponding V-V bond lengths using line profile data along the 'a-b' plane of the HRTEM image data, as shown in Figure S5 and Table S1. The in-plane lattice parameter of the $(2 \times 2)$ CDW is found to be $a_{CDW} = 0.65 \pm 0.01$ nm, whereas for $(1 \times 1)$ it is $a_{Bragg} = 0.32 \pm 0.01$ nm. Figure S5 shows the HRTEM image of a three-layer-thick multilayer $VS_2$ along the interlayer (a-c) plane. The intensity line profile across the interlayer is shown in Figure S5. The interlayer distance ($c = 0.59 \pm 0.01$ nm) is found to be extended compared to the standard value of 0.57 nm (ICDD PDF # 01-089-1640), which supports the intercalation-induced interlayer expansion in multilayered $VS_2$ [48, 88]. Figures 5 (c) and (d) represent a schematic of the top and side views of the $(2 \times 2)$ superstructure with intercalated Vanadium atoms (as marked in red and denoted by $V_I$). The counts in the line profile data at these intercalated positions increase due to a higher effective atomic number for this positional overlap (A-A stacking of the pristine lattice atoms and intercalated atoms) when looking from the top surface or along the c-direction of incoming electron beams. As a result of this self-intercalation, we get $(2 \times 2)$ superlattice structure in the multilayer self-intercalated $VS_2$.

The presence of self-intercalated vanadium atoms should increase if we probe a larger number of interlayers in a multilayer self-intercalated $VS_2$ flake. Hence, we have further studied the presence and charge state of these self-intercalated vanadium atoms by angle- or depth-dependent XPS study. Figure 6(a) and (b) show the high-resolution photoelectron spectra for the V 2p and S 2p core-levels, respectively. The V2p core shell splits into four main peaks, assigned to $V^{3+}$ ($2p_{3/2}$), $V^{4+}$ ($2p_{3/2}$) and their spin-orbit counter parts $V^{3+}$ ($2p_{1/2}$), and $V^{4+}$ ($2p_{1/2}$). The fitted peak parameters are given in the supporting information (Table S4). Each of these individual peaks further deconvoluted into two peaks. $V^{4+}$($2p_{3/2}$), arising from Vanadium oxidation state $4^+$, is split into two peaks at 516.18 and 517.22 eV, attributed to the 1T and 2H polymorphs of $VS_2$, respectively. The small differences arise from minor variations in the chemical environment, stemming from the dissimilar V-S bond-lengths and the stacking order in the two polymorphs. The $V^{3+}$ ($2p_{3/2}$), arising from Vanadium oxidation state $3^+$, is deconvoluted into two peaks at 512.95 and 513.99 eV, assigned to $V_2(SO_4)_3$ and $V_5S_8$, respectively. A small $V_2(SO_4)_3$ is a signature of small surface oxidation on ex-situ measured samples coming from the adsorption of environmental oxygen species with the dangling bonds at the sulfur vacancy sites, and $V_5S_8$ is due to the vanadium self-intercalation in $VS_2$ layers [48]. The S 2p core shell splits into $S^{2-}$ ($2p_{3/2}$), $S^0$ ($2p_{3/2}$), $S^{6+}$ ($2p_{3/2}$) and their spin orbit counter parts $S^{2-}$ ($2p_{1/2}$), $S^0$ ($2p_{1/2}$), $S^{6+}$ ($2p_{1/2}$) peaks. We have shown $S^{6+}$ ($2p_{3/2}$) and $S^{6+}$ ($2p_{1/2}$) as a single peak in the fit of S 2p. The $S^{2-}$ ($2p_{3/2}$) is coming from the metal sulfide bond of $VS_2$, $S^0$ ($2p_{3/2}$) is coming from the small atomic sulfur species on the flake which may arise from the bulk precursor $VS_2$ preparation during cooling of the chemical vapor deposition (CVD) furnace after vacuum drying or annealing at Ar atmosphere, and $S^{6+}$ ($2p_{3/2}$) is coming from the small adsorbed oxygen species at the dangling bonds of sulfur vacancy sites and forming small $V_2(SO_4)_3$ on the surface of multilayer $VS_2$ during the ex situ sample preparation. A depth-dependent study was performed by varying the emission angle, as shown in the schematic of Figure 6(c). The variation in electron escape depth ($d$) is a function of the angle of the analyzer ($\theta$) and follows the relation $d \approx \lambda_e \sin\theta$, where $\lambda_e$ is the inelastic mean free path of the electron ejected from the V 2p orbital for a given kinetic energy. Intensities of all of the spectra are normalized for accurate comparison. It can be seen in Figure 6(d) that increasing $\theta$ (and therefore $d$) leads to enhanced contributions from $V^{3+}$ ($2p_{3/2}$). The areas of the curves for the peaks due to $V^{4+}$ ($2p_{3/2}$) remain almost constant. The probing depths $d$ were calculated to be 46.1 (7 Layers), 56.5 (9 Layers), and 63.1 (10 Layers) Å, corresponding to the analyzer angles at 45, 60, and 75°, respectively. The calculated area ratio [$V^{3+}$($2p_{3/2}$) / $V^{4+}$ ($2p_{3/2}$)] increases significantly as a function of probe depth, as shown in Figure 6(e). From the data presented above, Figure S7 and Table S5 (Supporting Information), it can be inferred that there is direct evidence of $V^{3+}$ self-intercalation, which increases as we move from the surface to the bulk.

The self-intercalation of $V^{3+}$ ions may be triggered by sulfur vacancies (or loss) during annealing, as observed in other TMDCs as well [48, 89]. When the intercalated $V^{3+}$ ions is less than $(1/4)^{th}$ of the $V^{4+}$ ions present, the resulting stoichiometric phases are $V_5S_{8+x}$ and the resultant superstructure is $(2\times2)$ [33, 89]. If the intercalants reside at the octahedral voids of the 1T-$VS_2$, then the resulting superstructure becomes commensurate (i.e. $2 \times 2$) and if the intercalants reside at the tetrahedral voids or tilted octahedral voids of the 1T-$VS_2$ crystal structure, then the resulting superstructure becomes incommensurate (e.g. $(\sqrt{3} \times \sqrt{3})$ R 30°) [33]. From this, we can conclude that in our case, the resulting $V^{3+}$ intercalants reside at the octahedral voids in the 1T-$VS_2$ crystal lattice. Incomplete migration of intercalated metal ions creates a glassy phase, which can be detected in the electron diffraction pattern with shadow rings outside the Bragg spots and the central beam [33]. As we have not detected such ED patterns in our sample, the intercalation is likely complete in the octahedral void sites to make a commensurate $(2 \times 2)$ superstructure. The self-intercalation of $V^{3+}$ found in the 1T-$VS_2$ multilayers indicates [87, 38] a strong $V^{3+}$-$V^{4+}$ interlayer coupling mediated by Sulfur. The strong interlayer coupling between the CDW plane of V $d^2$ orbitals ($V^{3+}$) of superlattices and V $d^1$ orbitals ($V^{4+}$) of Bragg lattices in the superstructure $(2 \times 2)$ results in transfer of (1/3) electron per unit supercell from the CDW V-sites to the Bragg V-sites, resulting in a strong interlayer V-S-V coupling and making this superstructure more stable. This partial electron transfer of (1/6) electrons from $V^{3+}$ to $V^{4+}$ in the downward direction and (1/6) electrons from $V^{3+}$ to $V^{4+}$ in the upward direction causes the final oxidation state of lattice V to $V^{3.92+}$ from $V^{4+}$ and the oxidation state of intercalated V to $V^{3.33+}$[Figure S8 (Supporting Information)]. The CDW superstructure observed in the ML changes to (2×2) order as we increase the thickness from ML to multilayers due to self-intercalation of $V^{3+}$ ions, resulting charge transfer ($e^-$ self-doping), and strain. This type of interlayer charge transfer and electron doping may reduce the in-plane d-orbital hybridization and destroy ML incommensurate CDW superstructures [32, 6]. However, commensurate CDWs are found to be stable upon $e^-$ doping [85], and as a consequence, we get a stable commensurate $(2 \times 2)$ CDW order, and it is also possible to get both the ML commensurate $(\sqrt{7} \times \sqrt{7})$ R 19.1° and $(4 \times 4)$ CDW orders in multilayers of $VS_2$ [51, 83].

## 4. Conclusions

Our research presents a significant advancement in the experimental detection of the missing novel commensurate $(\sqrt{7} \times \sqrt{7})$ R 19.1° CDW order in the quasi-2D layered transition metal dichalcogenide (TMDC) prototype 1T-$VS_2$. The hexagonal crystal structure features a clustering of six nearest-neighbor metal atoms around a central metal atom, creating the CDW superstructure. Our results demonstrate that as the lattice distortion in 1T-$VS_2$ varies from 2.51% to 5.3%, the monolayer (ML) charge density wave transitions from $(4\times4)$ to $(\sqrt{7} \times \sqrt{7})$ R 19.1°. The transition of CDW orders among competing phases is facilitated by the existence of corresponding soft-phonon modes and similar energy reductions of $-11.16$ $meV$ to $-12.57$ $meV$ compared to the pristine undistorted $(1 \times 1)$ 1T-$VS_2$ structure. This study examines three critical issues: first, the characteristics and origin of the novel competing CDW orders in a monolayer of 1T-$VS_2$ at room temperature; second, the impact of bilayers or small-angle Moiré patterns on the stability of our monolayer $(\sqrt{7}\times\sqrt{7})$ R 19.1° CDW; and third, the dimensional crossover of $(\sqrt{7}\times\sqrt{7})$ R 19.1° CDW to a new type-III $(2 \times 2)$ CDW order in multilayer $VS_2$. Our findings present promising opportunities for investigating and regulating charge-density-wave superstructure orders with varying layer counts in $1T-VS_2$ and other two-dimensional charge-density-wave materials synthesized via liquid-phase exfoliation and subsequent stamp-transfer process (LPESTP).

## 5. Supporting Information Available

## 6. Acknowledgements

PP acknowledges the ANRF research grant (ARG/2025/001743/ENS) for financial support. AKR acknowledges the Science and Engineering Research Board (SERB) for financial support in the form of a Distinguished Fellowship (No. SB / DF / 008/2019) and additional support through the Indian National Science Academy (INSA) Senior Scientist Program. SP thanks the Council of Scientific and Industrial Research (CSIR), India, for the CSIR-NET Fellowship. KC thanks the Department of Science and Technology (DST), India, for the INSPIRE Fellowship (IF210326). The authors thank Mr. Tukai Singha of Saha Institute of Nuclear Physics (SINP), Kolkata, India, for his help with TEM measurements, and Mr. Subhadip Sardar at CSIR-CGCRI for his help with simulating Moiré patterns. JS and MK acknowledge the National Supercomputing Mission (NSM) (supported by DST and MeitY) for providing computing resources of 'PARAM RUDRA' at S.N. Bose National Centre for Basic Sciences, Kolkata, India.

## CRediT authorship contribution statement

**Samanta Pal:** Writing- Original Draft, Conceptualization, Methodology, Investigation, Data Curation, Formal Analysis, Validation, Visualization, Software, Writing- Review and editing. **Kaustuv Chatterjee:** Data Curation, Investigation, Formal Analysis, Visualization, Software, Writing- Review and editing. **Jyotirmoy Sau:** Data Curation, Investigation, Formal Analysis, Validation, Software, Writing- Review and editing. **Biswarup Satpati:** Data Curation, Investigation, Validation. **Manoranjan Kumar:** Supervision, Investigation, Validation, Resources. **Arup Kumar Raychaudhuri:** Writing- Review and editing, Visualization, Investigation, Validation, Supervision, Funding Acquisition, Project Administration, Resources. **Prabir Pal:** Writing- Review and editing, Conceptualization, Investigation, Visualization, Validation, Supervision, Funding Acquisition, Project Administration, Resources.

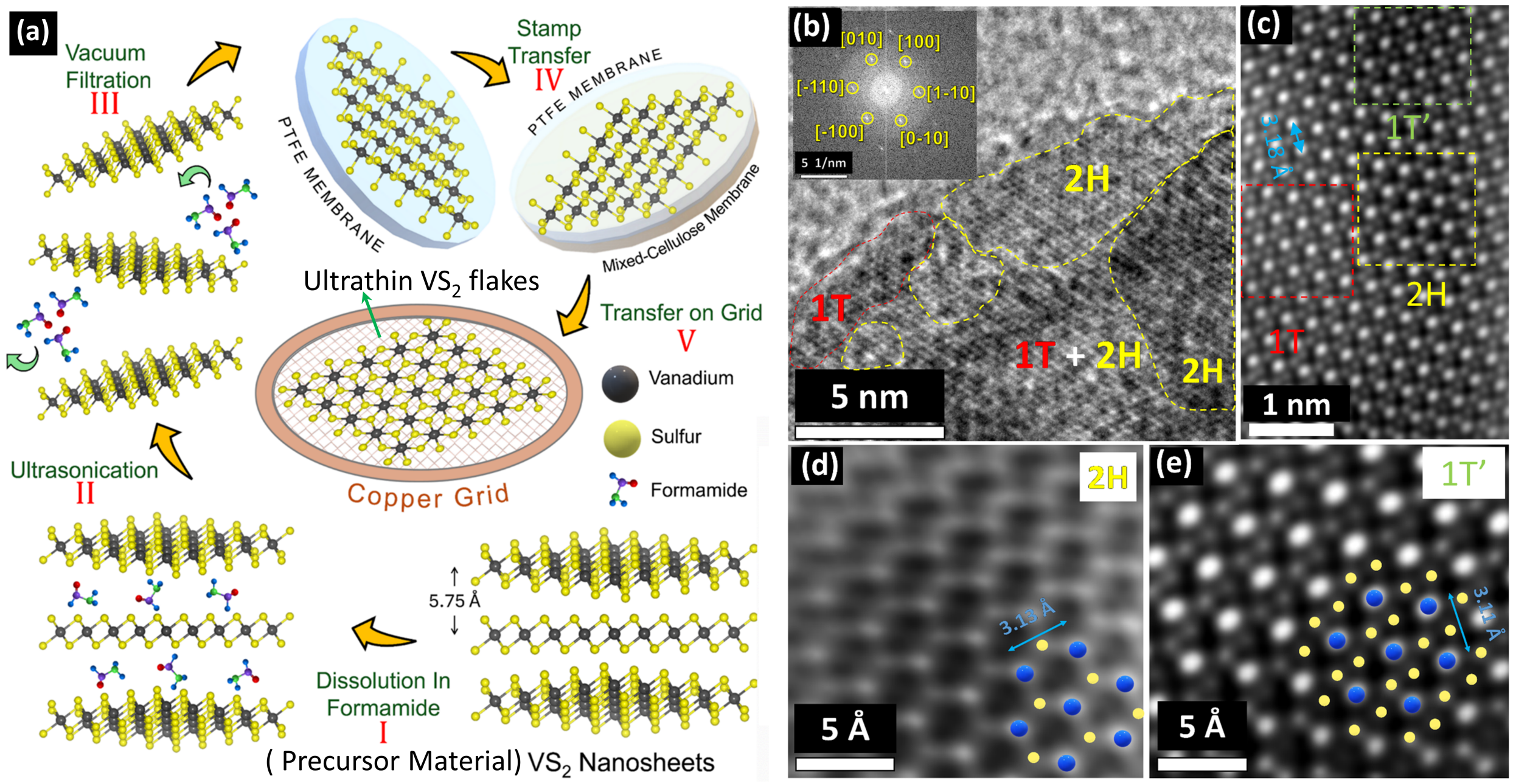

**Figure 1:** Co-existence of polymorphic phases (1T, 2H and 1T$^{'}$) in ultrathin $VS_2$ flakes. (a) shows a schematic of the ultrathin layer transfer method by the LPESTP, (b) High-resolution transmission electron microscopy (HRTEM) image of an ML $VS_2$ flake at 295 K in real space, which shows the co-existence of 1T and 2H polymorphs as marked. Inset is the fast Fourier transform (FFT) of the above real-space HRTEM showing the reciprocal-space spots along the [001] zone axis and its hexagonal symmetry. (c) HRTEM image of a portion of the flake showing different regions of polymorphic phase transitions as marked. Vanadium (larger and brighter spots) and sulfur (smaller and less bright spots) atoms are clearly visible from the atomically resolved HRTEM image. (d) 2H and (e) 1T$^{'}$ polymorphs are identified from the atomically resolved HRTEM image. Atomic structure models of the two polymorphic phases are shown, which confirm their co-existence in ML $VS_2$ flake.

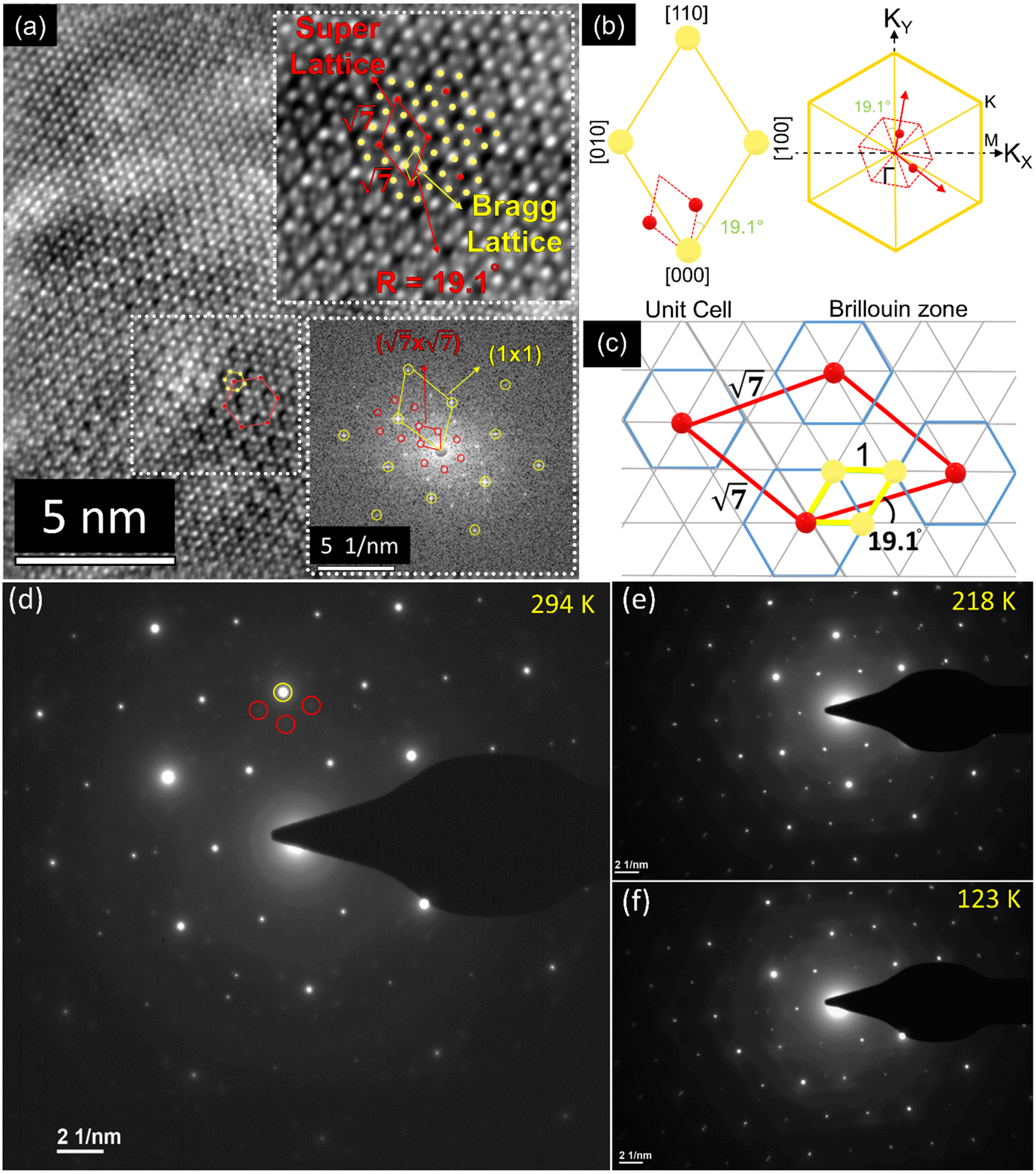

**Figure 2:** Direct observation of a novel CDW order in ML 1T-$VS_2$. (a) HRTEM image of an ML of 1T-$VS_2$ flake taken at 295 K. A clearer view of the pristine Bragg lattice (marked yellow) and CDW superlattice (marked red) is shown in the top-right corner inset. The bottom-right inset shows the fast Fourier transform (FFT) of the real-space HRTEM image, revealing distinct Bragg and CDW-order spots in reciprocal space, marked by yellow and red circles, respectively. (b) Schematics of the unit cells and first Brillouin zones for pristine $(1 \times 1)$ and $(\sqrt{7} \times \sqrt{7})$ CDW order, as marked by yellow and red, are shown, respectively, along with the CDW modulation positions (red points). (c) Schematic illustration of the $(\sqrt{7} \times \sqrt{7})$ R 19.1° CDW (marked by red) and corresponding undistorted $(1 \times 1)$ Bragg lattice (marked by yellow) unit cells showing the relative angle and surface re-arrangements. (d), (e) and (f) show the SAED patterns of such a CDW layer at temperatures 294 K, 218 K, and 123 K, respectively.

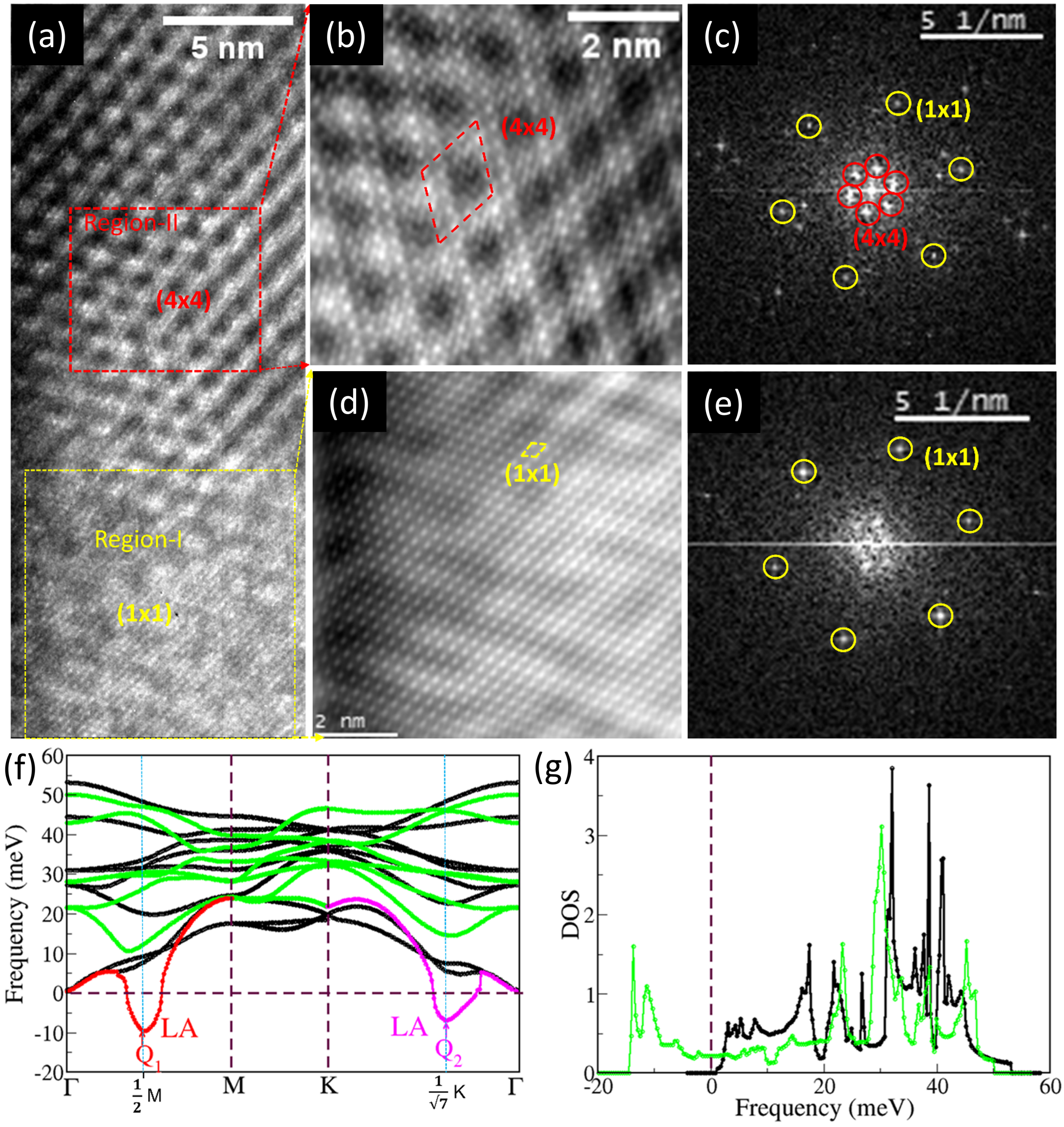

**Figure 3:** (a) Displays the HRTEM image of a $VS_2$ flake, featuring both the pristine $(1 \times 1)$ structure (Region-I) and the $(4{\times}4)$ CDW superstructure (Region-II). Panels (b) and (c) present a zoomed-in inverse fast Fourier transform (IFFT) view and fast Fourier transform (FFT) of Region-II, respectively, while panels (d) and (e) provide similar views for Region-I. The unit cells of the $(1 \times 1)$ and $(4 \times 4)$ CDW superstructures are indicated by yellow and red rhombuses, respectively, and the Bragg and superstructure diffraction spots are denoted by yellow and red circles. (f) illustrates the phonon dispersion relations for monolayer $(1 \times 1)$ $1T$-$VS_2$ (black) alongside the reconstructed $(\sqrt{7} \times \sqrt{7})$ R 19.1° superstructure (green). Here, $Q_1$ and $Q_2$ represent the notable imaginary modes linked to lattice instabilities. The two softened modes of the $(\sqrt{7} \times \sqrt{7})$ R 19.1° CDW superstructure are highlighted in pink and red. (g) shows the corresponding phonon total density of states (DOS) for monolayer $1T$-$VS_2$ (black) and the $(\sqrt{7} \times \sqrt{7})$ R 19.1° superstructure (green), with negative frequencies illustrated for the $(\sqrt{7} \times \sqrt{7})$ R 19.1° superstructure to emphasize the presence of soft phonon modes.

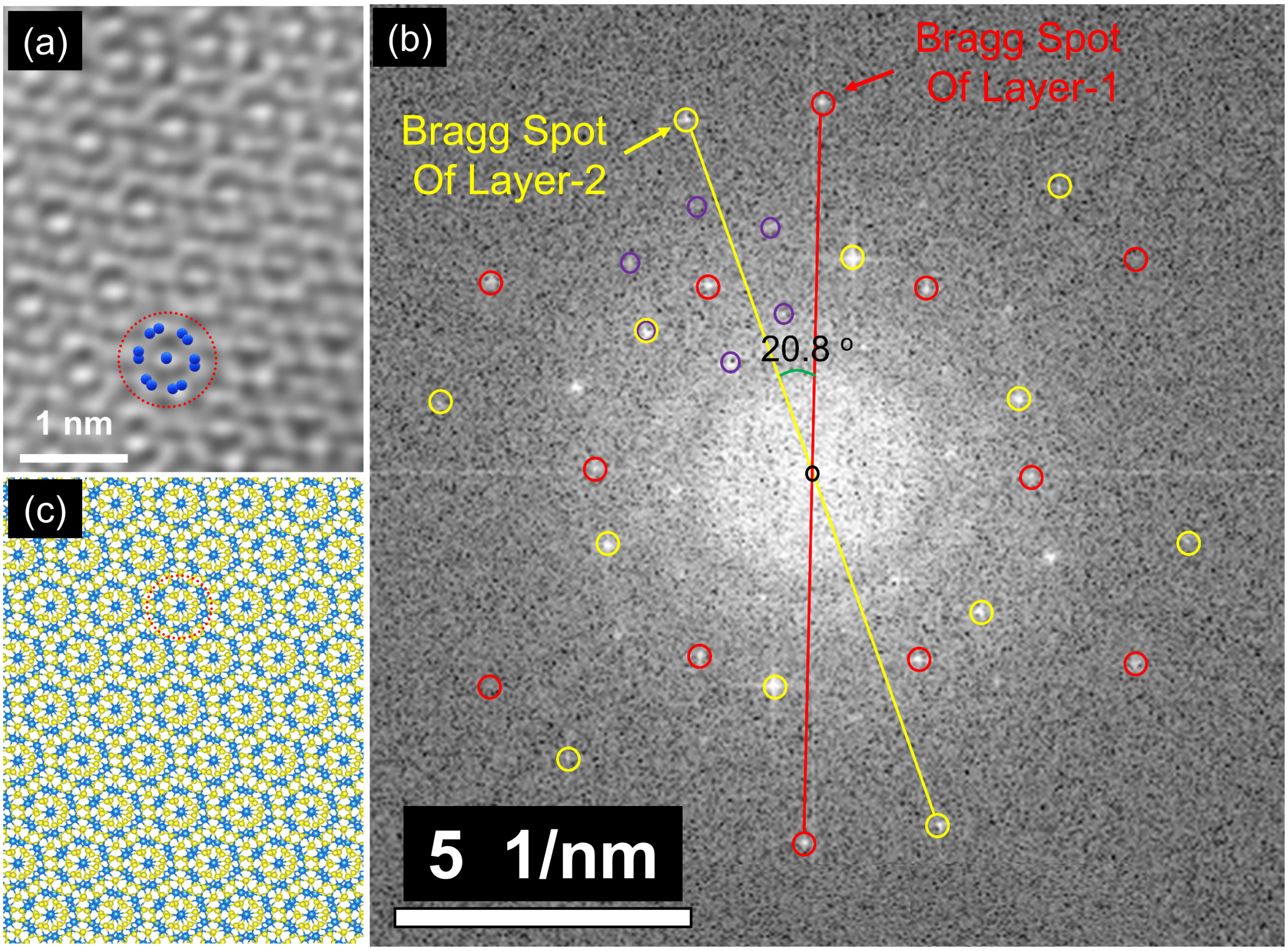

**Figure 4:** Moiré superlattice in a small-angle hetero-bilayer of the pristine $(1\times1)$ ML $1T-VS_2$ and $(\sqrt{7}\times\sqrt{7})$ R 19.1° ML CDW. (a) HRTEM image of the Moiré superlattice in real space taken at room temperature (295 K), (b) FFT of the HRTEM image of Figure 4(a) shows the Moiré superlattice spots in reciprocal space. The twist angle between the pristine $(1\times1)$ ML $1T-VS_2$ and $(\sqrt{7}\times\sqrt{7})$ R 19.1° ML CDW is estimated to be (20.8 - 19.1)° = 1.7°. (c) Simulated Moiré pattern of the twisted hetero-bilayer of pristine $(1\times1)$ $1T-VS_2$ and $(\sqrt{7}\times\sqrt{7})$ R 19.1° ML CDW superstructure well matches the observed pattern as in Figure 4(a).

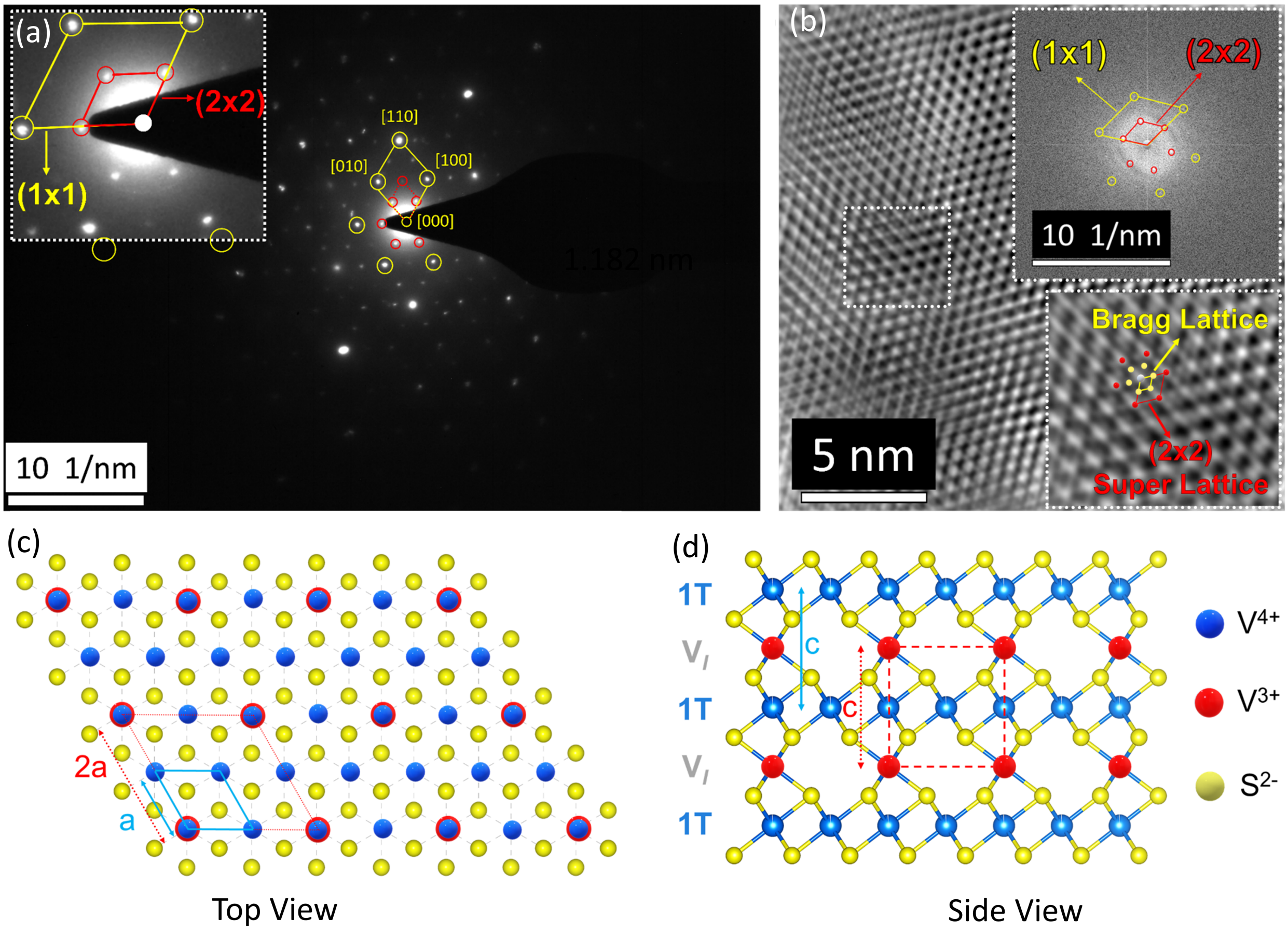

**Figure 5:** CDW superstructure in multilayer $VS_2$ flakes. (a) Electron diffraction (ED) pattern from a multilayer $VS_2$ flake showing sharp and intense hexagonally arranged pristine Bragg spots (marked in yellow) as well as sharp but less intense hexagonally arranged superlattice spots (marked in red). The inset shows a closer view of the hexagonally arranged diffraction spots for the pristine (yellow) and superlattice (red) structure near the central beam spot. The reciprocal-space unit cells of a $(1\times1)$ pristine Bragg and $(2\times2)$ superlattice are marked by yellow and red rhomboses. (b) HRTEM image of a multilayer $VS_2$ flake showing a clear $(2\times2)$ superlattice order. The top-right inset shows the FFT spots corresponding to the HRTEM image, clearly indicating the $(2\times2)$ superlattice spots and Bragg spots, as marked by red and yellow circles, respectively. The bottom-right inset is a zoomed-in view of the region marked by the white border, where pristine Bragg lattice points (yellow) and superlattice points (red) are marked. (c) and (d) show the schematic of the top and side views of the $(2\times2)$ superstructure with intercalated Vanadium atoms (red).

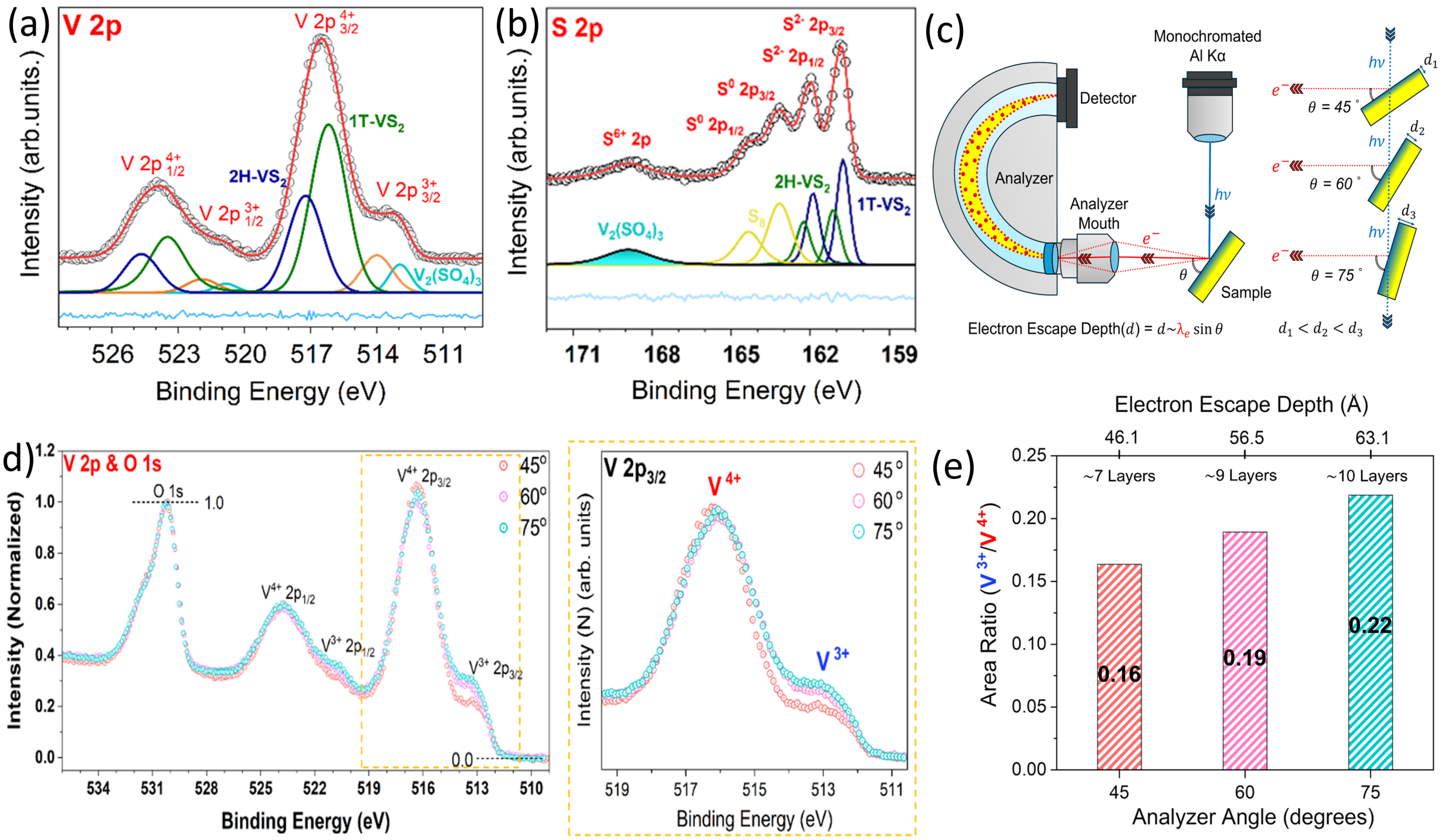

**Figure 6:** Angle (or electron escape-depth) dependent X-ray photoelectron spectroscopy (XPS) from multilayer $VS_2$ flakes. (a) and (b) show the high-resolution V 2p and S 2p core-level XPS spectra, respectively, at an angle of 45° between the sample top plane and the analyzer. The black circles indicate the experimental data, while the red line represents the resultant of the fitted curve, with individual peaks shown below, and the light blue line illustrates the magnitude of variation. The Shirley background is subtracted before fitting the curves. (c) shows the schematic of the angle-dependent measurement setup. The angle between the sample top surface and analyzer was varied by applying different tilts to the sample platen. (d) shows the normalized spectra of V 2p and O 1s region taken at different acquisition angles. An enlarged view of the V $2p_{3/2}$ region is shown with a marked yellow dashed rectangular area to highlight the change in area for the $V^{3+}$ component with electron escape depth or layer thickness. (e) shows the plot between the analyzer angle (or electron escape depth) and the area ratio of ($V^{3+}/V^{4+}$).

# Supporting Information

# Direct Observation of Novel Charge Density Wave Superstructures at Room Temperature in Ultrathin 1T-$VS_2$ Flakes

Samanta Pal,[†] Kaustuv Chatterjee,[†] Jyotirmoy Sau,[¶] Biswarup Satpati,[§] Manoranjan Kumar,[‖] A. K. Raychaudhuri,[*,⊥] and Prabir Pal[*,†]

†CSIR Central Glass and Ceramic Research Institute, 196 Raja S.C. Mullick Road, Kolkata-700032, India

‡Academy of Scientific and Innovative Research (AcSIR), Ghaziabad, 201002, India

¶Department of Physics and Astronomy, Uppsala University, Box-516, S-75120 Uppsala, Sweden

§Surface Physics and Material Science Division, Saha Institute of Nuclear Physics, 1/AF, Bidhannagar, Kolkata 700 064, INDIA.

‖Department of Condensed Matter Physics and Materials Science, S.N. Bose National Centre for Basic Sciences, JD Block, Sector III, Salt Lake, Kolkata-700106, India

⊥UGC-DAE Consortium for Scientific Research, Kolkata Centre, Sector III, Bidhan Nagar, Kolkata 700106, India

E-mail: arupraychaudhuri4217@googlemail.com; prabir.cgcri@csir.res.in

## 1. Growth and characterization of bulk precursor $VS_2$ sample

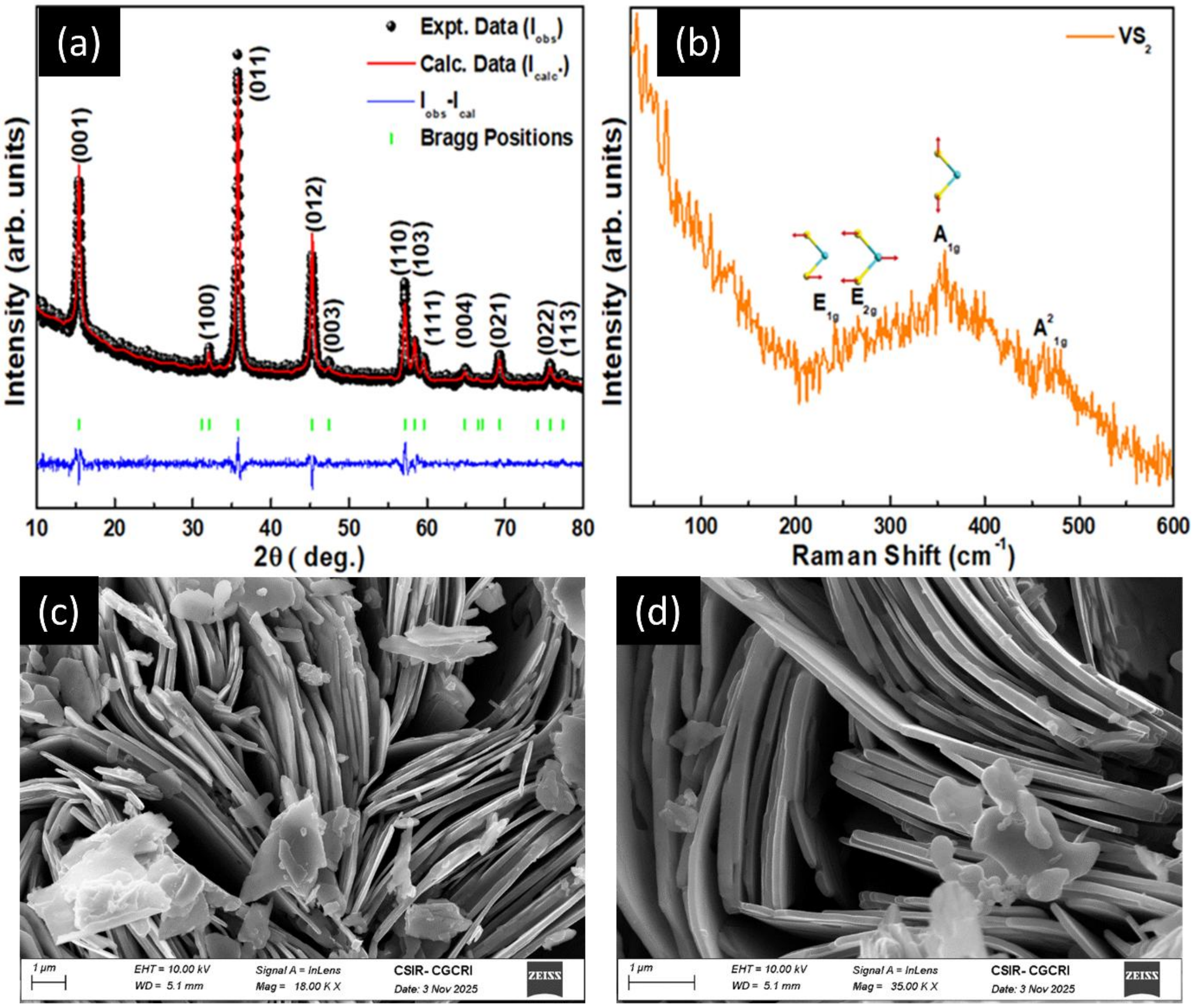

**FIGURE S1:** (a) X-ray diffractograms of the bulk precursor $VS_2$ and its Rietveld refinement obtained lattice parameters a=b=3.22 Å and c=5.74 Å. (b) Raman spectra of bulk precursor $VS_2$ with its characteristic vibrational modes marked. (c) and (d) FESEM images of the bulk precursor $VS_2$ showing the lateral sizes (6-8 μm) of the bulk precursor $VS_2$ flakes and its thickness in the range of 100 -150 nm.

## 2. Characterization of LPESTP-obtained multilayer $VS_2$ flakes

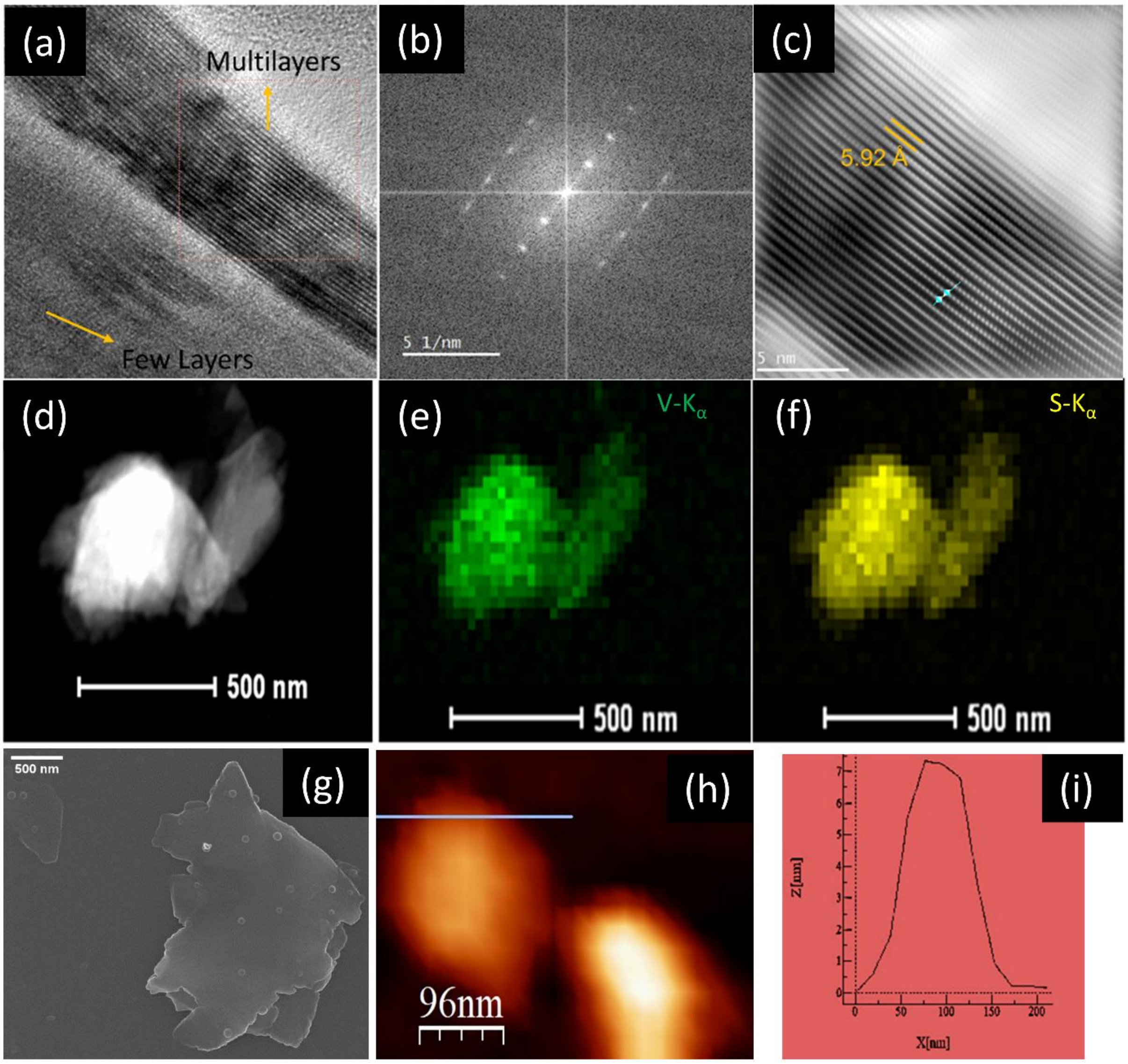

**FIGURE S2**: (a) HRTEM image of multilayer $VS_2$ flakes showing its layered structure. (b) FFT of the HRTEM image of (a) showing linear spots, which are commonly seen in layered materials. (c) shows the Fourier filtered image of (a) showing extended interlayer distance (5.92 Å) for $VS_2$ and the mark from where the line profile for the distance measurement is taken. (d)-(f) show the STEM HAADF image of a multilayer $VS_2$ flake and its elemental mapping for V-Kα and S-Kα, respectively. (g) shows the FESEM image of a multilayer $VS_2$ flake with a lateral dimension of approximately 2-2.5 µm. (h) shows the atomic force micrograph (AFM) topography image of a multilayer thick $VS_2$ flake with a lateral size of around 200 nm, and (i) shows the height profile along a line from the flake as marked in (h) with a height of around 7 nm (~12 layers).

## 3. Thickness and lateral sizes of the $VS_2$ flakes transferred onto the $SiO_2$/Si substrate by LPESTP

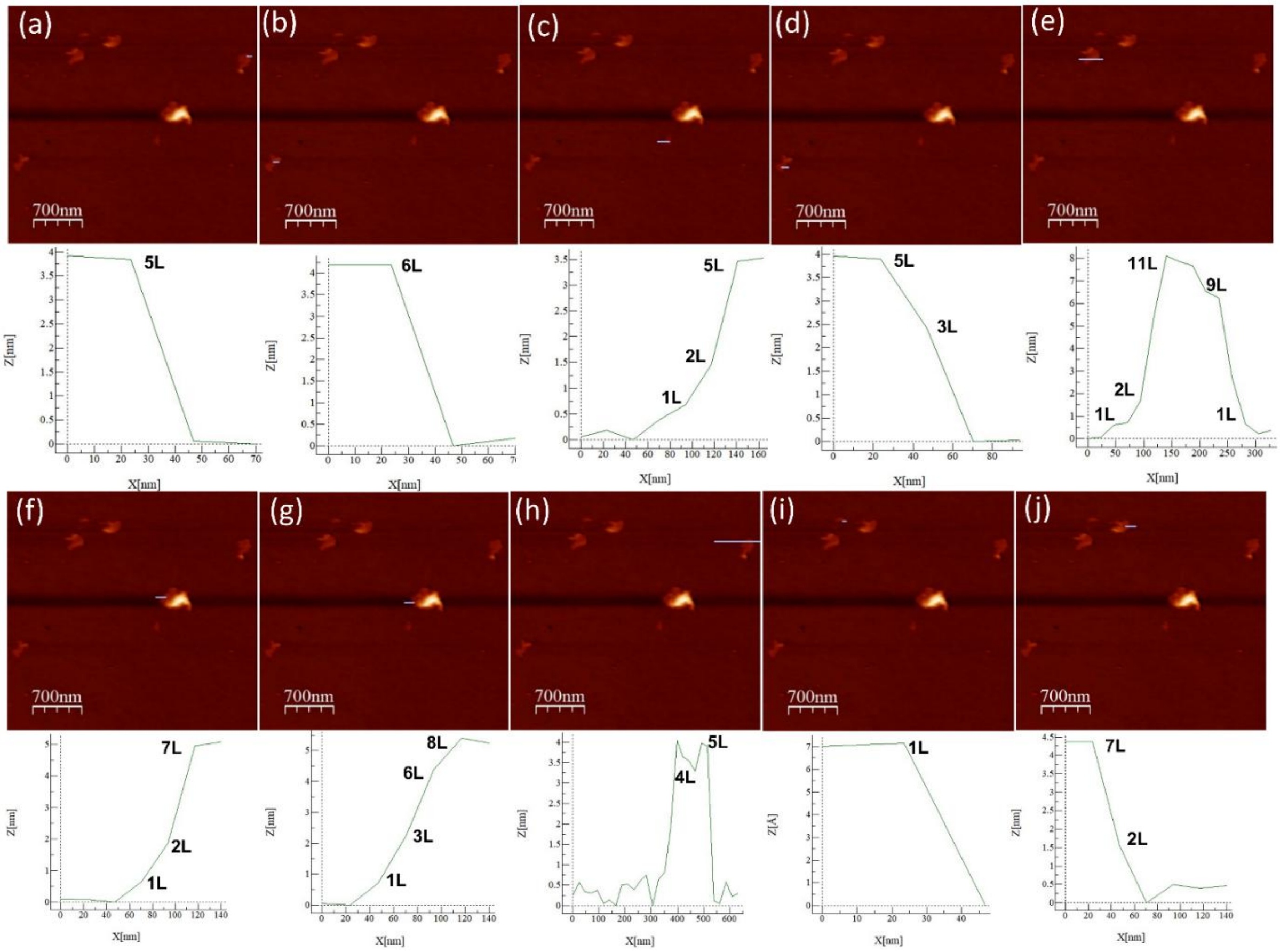

**FIGURE S3**: (a-j) AFM topography images and corresponding thickness profiles of stamp-transferred $VS_2$ flakes on $SiO_2$/Si substrates. The flakes are marked by small white lines from where the thickness profiles are taken. The measured thicknesses of the flakes are shown just below each topography image. We can see from the AFM height profiles that LPESTP obtained $VS_2$ flakes have lateral (L) sizes $0.2 \leq L \leq 0.5$ µm and thickness (t) distribution in the range $0.7 \leq t \leq 8$ nm, which correspond to 1-11 layers of $VS_2$. The monolayer $VS_2$ thickness is found to be ~ 0.7 nm, which matches well with the literature.

## 4. V-V Bond lengths in ( $\sqrt{7} \times \sqrt{7}$ ) R19.1° CDW superstructure in monolayer

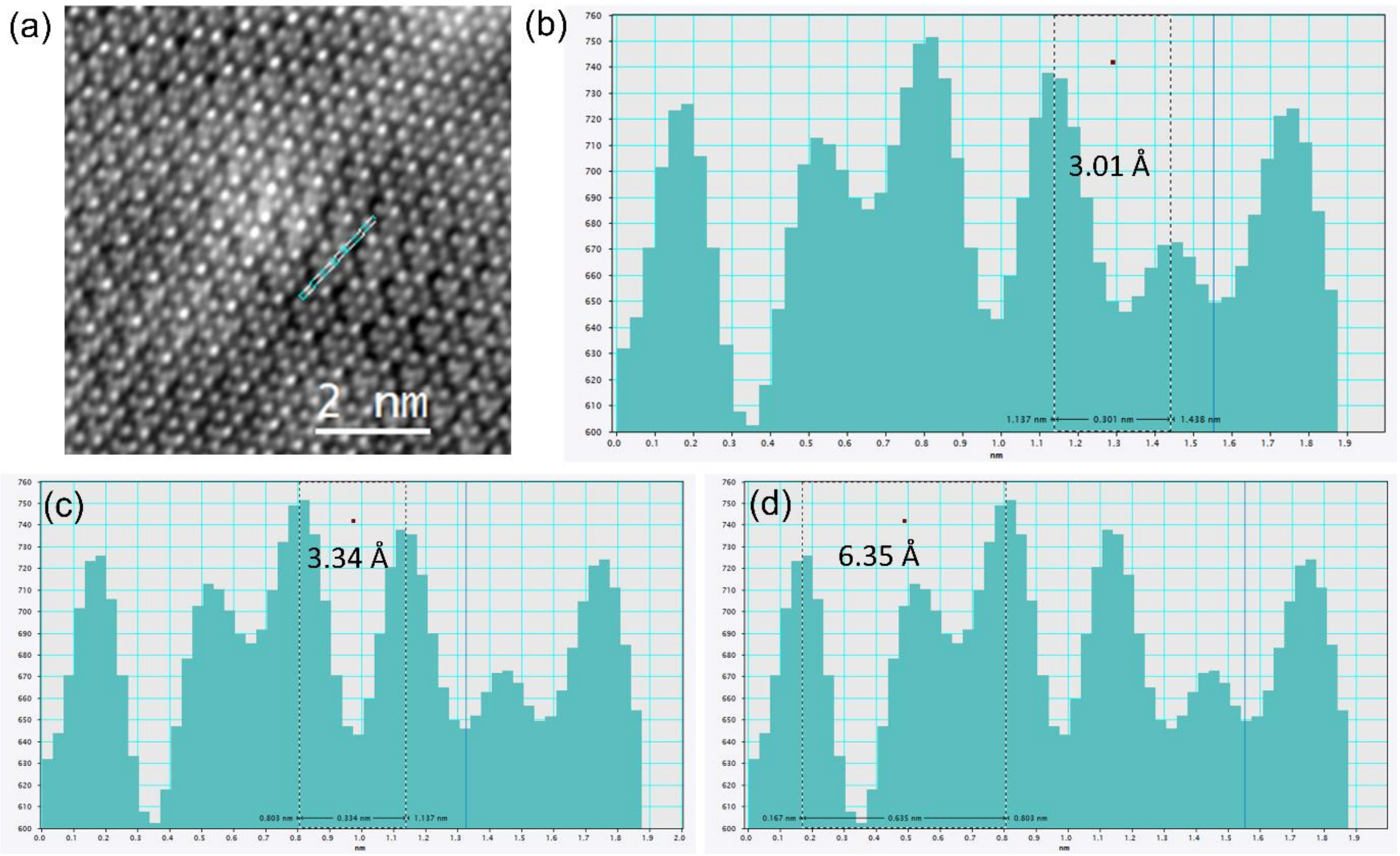

**FIGURE S4**: (a) HRTEM image of ($\sqrt{7} \times \sqrt{7}$) R19.1° CDW superstructure. (b) and (c) are distances between the two types of nearest neighbour V atoms. These two types of V-V bond lengths found in the CDW phase are consistent with the study by **Sun** *et al*. [1]. (d) Distance between the second nearest neighbour V atoms.

## 5. V-V Bond lengths in (2 × 2) CDW superstructure in multilayers

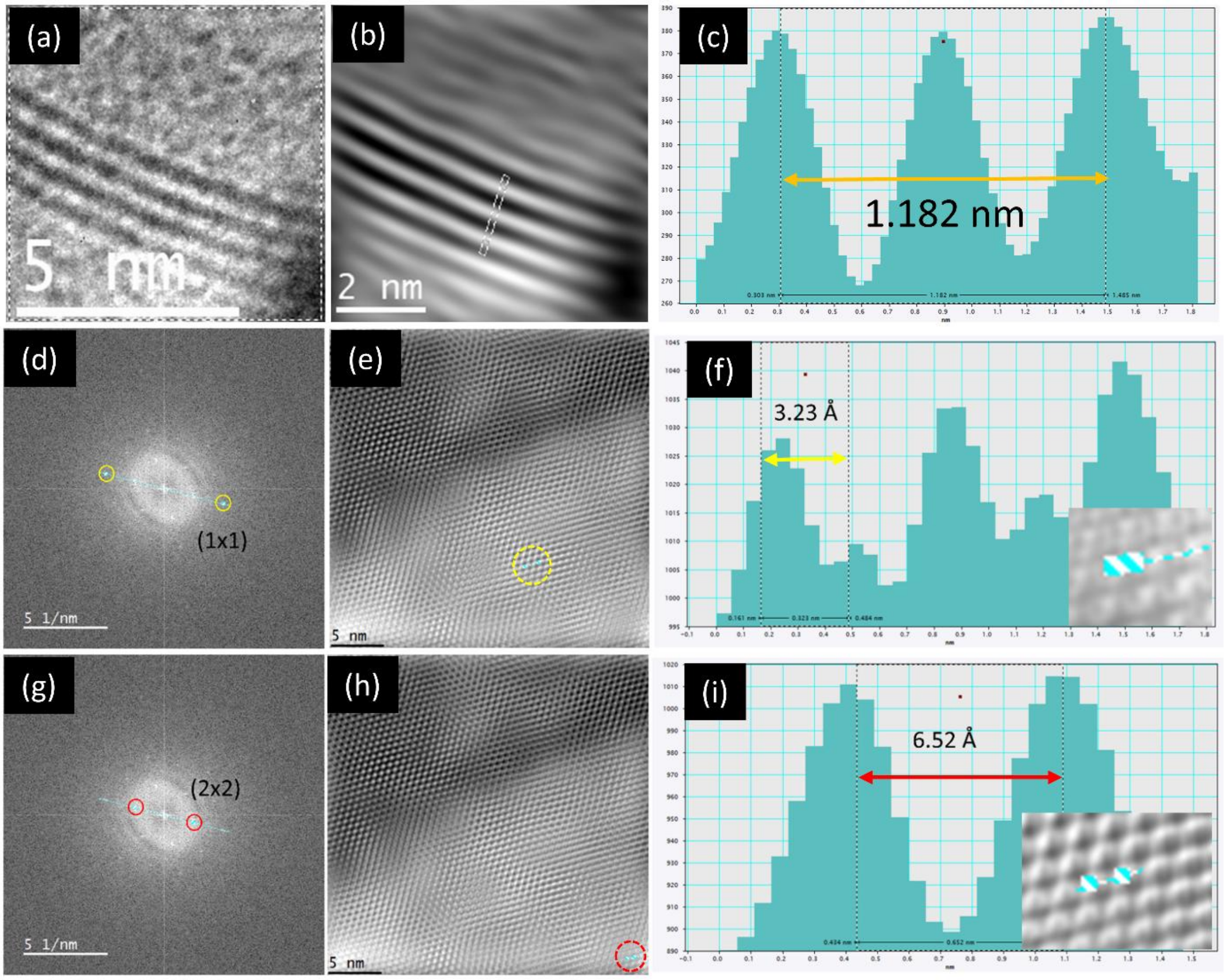

**FIGURE S5**: (a), (b), and (c) represent, respectively, the TEM image, its IFFT image, and the interlayer distance between the two alternate layers of a four-layer thick $VS_2$ flake. The calculated interlayer distance is (1.182/2) = 0.591 nm. (d), (e), and (f) are, respectively, the FFT, IFFT, and nearest V-V bond lengths for the Bragg lattice, while (g), (h), and (i) are for the (2 × 2) CDW lattice. In the inset of (f) and (i), the lines are shown along which the V-V line profile is drawn.

## 6. Moiré superlattice in twisted hetero-bilayers of 1T-$VS_2$

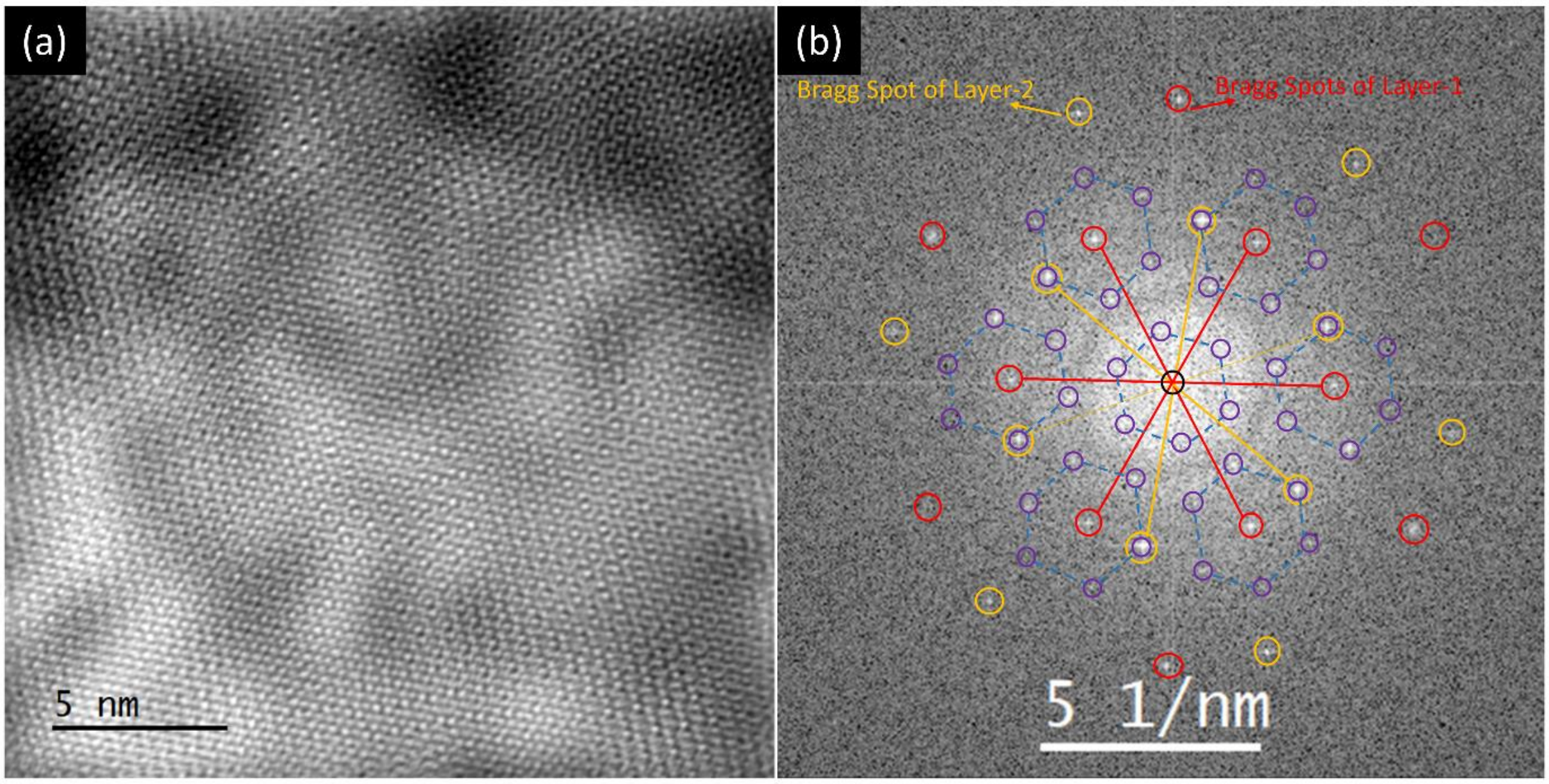

**F I G U R E  S 6**: (a) Moiré superlattice patterns in the twisted hetero-bilayer of a pristine (1 × 1) and ($\sqrt{7} \times \sqrt{7}$) R19.1° CDW layer of 1T-$VS_2$ at a twist angle of 1.7°. (b) FFT of (a), showing clearly the Bragg spots of the two layers making a (19.1+1.7) ° = 20.8° angle with each other as marked by red and yellow circles, respectively. The Moiré periodicity spots are marked by violet circles.

## 7. Measured bond-lengths in polymorphic and CDW phases of $VS_2$

| Sl. No. | Phases | Bond Lengths (Å) | |
|---|---|---|---|
| 1. | 1T | 3.18 (V-V) | 2.09 (V-S) |
| 2. | 2H | 3.13 (V-V) | 2.13 (V-S) |
| 3. | 1T' | 3.11 (V-V) | |
| 4. | $\sqrt{7} \times \sqrt{7}$ | 3.34 $(V\text{-}V)_{long}$<br>3.01 $(V\text{-}V)_{short}$<br>PLD = 5.3 %<br><br>8.29 (V-V) (cdw periodicity) | |
| 5. | 4 × 4 | (a)<br>3.24 $(V\text{-}V)_{long}$<br>3.10 $(V\text{-}V)_{short}$<br>PLD = 2.51 %<br><br>(b)<br>3. 32 $(V\text{-}V)_{long}$<br>3.07 $(V\text{-}V)_{short}$<br>PLD = 3.46 %<br><br>12.68 (V-V) (cdw periodicity) | |
| 6. | 2 × 2 | 3.23 $(V\text{-}V)_{long}$<br>6.41 (V-V) (cdw periodicity) | |

**Table-S1**: Measured V-V and V-S bond lengths from HRTEM data of different polymorphic phases and CDW orders in $VS_2$. The uncertainty in the bond-length measurement is around 0.02 Å.

# 8. CDW phases observed in a monolayer of different TMDCs

| Sl. No | Material | Superstructures | Temperature | Ref. |
|---|---|---|---|---|
| 1. | $TiSe_2$ | $2 \times 2$ | 232 K | [2] |
| 2. | $TiSe_2$ | $2 \times 2$ | 200 K | [3] |
| 3. | $TiTe_2$ | $2 \times 2$ | 92 K | [4] |
| 4. | $TaSe_2$ | $3 \times 3$ | 5 K | [5] |
| 5. | $VSe_2$ | $(\sqrt{3} \times \sqrt{7})$, $(\sqrt{3} \times 2)$ | 79 K | [6] |
| 6. | $VSe_2$ | $\sqrt{3}$ R 30° × $\sqrt{7}$ R 19.1° | 80 K<br>100 K | [7] |
| 7. | $VS_2$ | 9 × $\sqrt{3}$ R 30° | 300 K | [8] |
| 8. | $VS_2$ | $\sqrt{21}$ R 10.9° × $\sqrt{3}$ R 30° | 4.5 K | [9] |
| 9. | $V_9S_{16}$,<br>$V_5S_{8+x}$ | $(\sqrt{3} \times \sqrt{3})$ R 30° | 110 K | [10] |
| 10. | $VS_2$ | $(\sqrt{7} \times \sqrt{7})$ R 19.1°,<br>4 × 4 | 295 K | This Work |

**Table-S2**: A comparative table showing superstructure orders for different ML TMDC materials with their measured temperatures.

## 9. Superstructures observed in metal-rich/intercalated TMDC materials ($M_{1+x}X_2$, where, M = V, Nb, Ta, Ti, Cr and X = S, Se, Te)

| Sl. No | Material | Superstructures | Temperature (K) | Ref. |
|---|---|---|---|---|
| 1. | $V_9S_{16}$<br><br>$V_5S_{8+x}$ | $(\sqrt{3} \times \sqrt{3})$ R 30° and (2×2)<br><br>$(\sqrt{3} \times \sqrt{3})$ R 30° and (2×2) | 110 K<br><br>110 K | [10] |
| 2. | $M_{1+x}Te_2$ (M = Ti, V, Cr)<br><br>$M_{1.25}Te_2$<br><br>$M_{1.33}Te_2$<br><br>$M_{1.50}Te_2$ | (2×2)<br><br>$(\sqrt{3} \times \sqrt{3})$ R 30°<br><br>(2×1) | RT and 20 K | [11] |
| 3. | $M_{1+x}Se_2$ (M = Ti, V)<br><br>$M_{1.25}Se_2$<br><br>$M_{1.33}Se_2$<br><br>$M_{1.50}Se_2$ | (2×2)<br><br>$(\sqrt{3} \times \sqrt{3})$ R 30°<br><br>(2×1) | 20 K | [12] |
| 4. | $V_{11}S_{16}$<br><br>$Ta_9S_{16}$<br><br>$Ta_7S_{12}$ | (2×2)<br><br>$(2 \times \sqrt{3}$ R 30°)<br><br>$(\sqrt{3} \times \sqrt{3})$ R 30° | RT | [13] |
| 5. | Multilayer $VS_2$ ($V_{1+x}S_2$, $0 < x < 0.1$) | (2×2) | RT | This work |

**Table-S3:** A comparative table showing superstructure orders for different metal-rich/intercalated TMDC materials.

# 10. XPS Fitting Parameters

| V 2p | | | | | | | | | | | | | | | |
|---|---|---|---|---|---|---|---|---|---|---|---|---|---|---|---|
| V $2p_{3/2}$ | | | | | | | | V $2p_{1/2}$ | | | | | | | |
| $V^{3+}$ | | | | $V^{4+}$ | | | | $V^{3+}$ | | | | $V^{4+}$ | | | |
| $V_2(SO_4)_3$ | | $V_5S_8$ | | 1T-$VS_2$ | | 2H-$VS_2$ | | $V_2(SO_4)_3$ | | $V_5S_8$ | | 1T-$VS_2$ | | 2H-$VS_2$ | |
| Peak Position (eV) | FWHM (eV) | Peak Position (eV) | FWHM (eV) | Peak Position (eV) | FWHM (eV) | Peak Position (eV) | FWHM (eV) | Peak Position [$\Delta_{SO}$] (eV) | FWHM (eV) | Peak Position [$\Delta_{SO}$] (eV) | FWHM (eV) | Peak Position [$\Delta_{SO}$] (eV) | FWHM (eV) | Peak Position [$\Delta_{SO}$] (eV) | FWHM (eV) |
| 512.95 | 1.15 | 513.99 | 1.64 | 516.18 | 1.92 | 517.22 | 1.84 | 520.83 [7.88] | 1.19 | 521.91 [7.92] | 1.67 | 523.47 [7.29] | 1.95 | 524.66 [7.44] | 1.83 |

| S 2p | | | | | | | | | | | | | |
|---|---|---|---|---|---|---|---|---|---|---|---|---|---|
| S $2p_{3/2}$ | | | | | | VS$2p_{1/2}$ | | | | | | $S^{6+}$ ($SO_4^{2-}$) | |
| $S^{2-}$ | | | | $S^0$ | | $S^{2-}$ | | | | $S^0$ | | Peak Position (eV) | FWHM (eV) |
| 1T-$VS_2$ | | 2H-$VS_2$ | | Peak Position (eV) | FWHM (eV) | 1T-$VS_2$ | | 2H-$VS_2$ | | Peak Position [$\Delta_{SO}$] (eV) | FWHM (eV) | | |
| Peak Position (eV) | FWHM (eV) | Peak Position (eV) | FWHM (eV) | | | Peak Position [$\Delta_{SO}$] (eV) | FWHM (eV) | Peak Position [$\Delta_{SO}$] (eV) | FWHM (eV) | | | | |
| 160.75 | 0.57 | 161.12 | 0.62 | 163.16 | 1.05 | 161.89 [1.14] | 0.60 | 162.25 [1.13] | 0.62 | 164.34 [1.18] | 1.15 | 168.90 | 2.09 |

| O 1s | | | |
|---|---|---|---|
| $O^{2-}$ | | $OH^-$ | |
| Peak Position (eV) | FWHM (eV) | Peak Position (eV) | FWHM (eV) |
| 530.41 | 1.36 | 531.79 | 1.44 |

**Table-S4**: XPS fitting parameters for V 2p, S 2p, and O 1s in our ex-situ prepared multilayer $VS_2$ sample.

## 11. Fitting of V 2p and O 1s core-level photoemission spectra at different emission angles

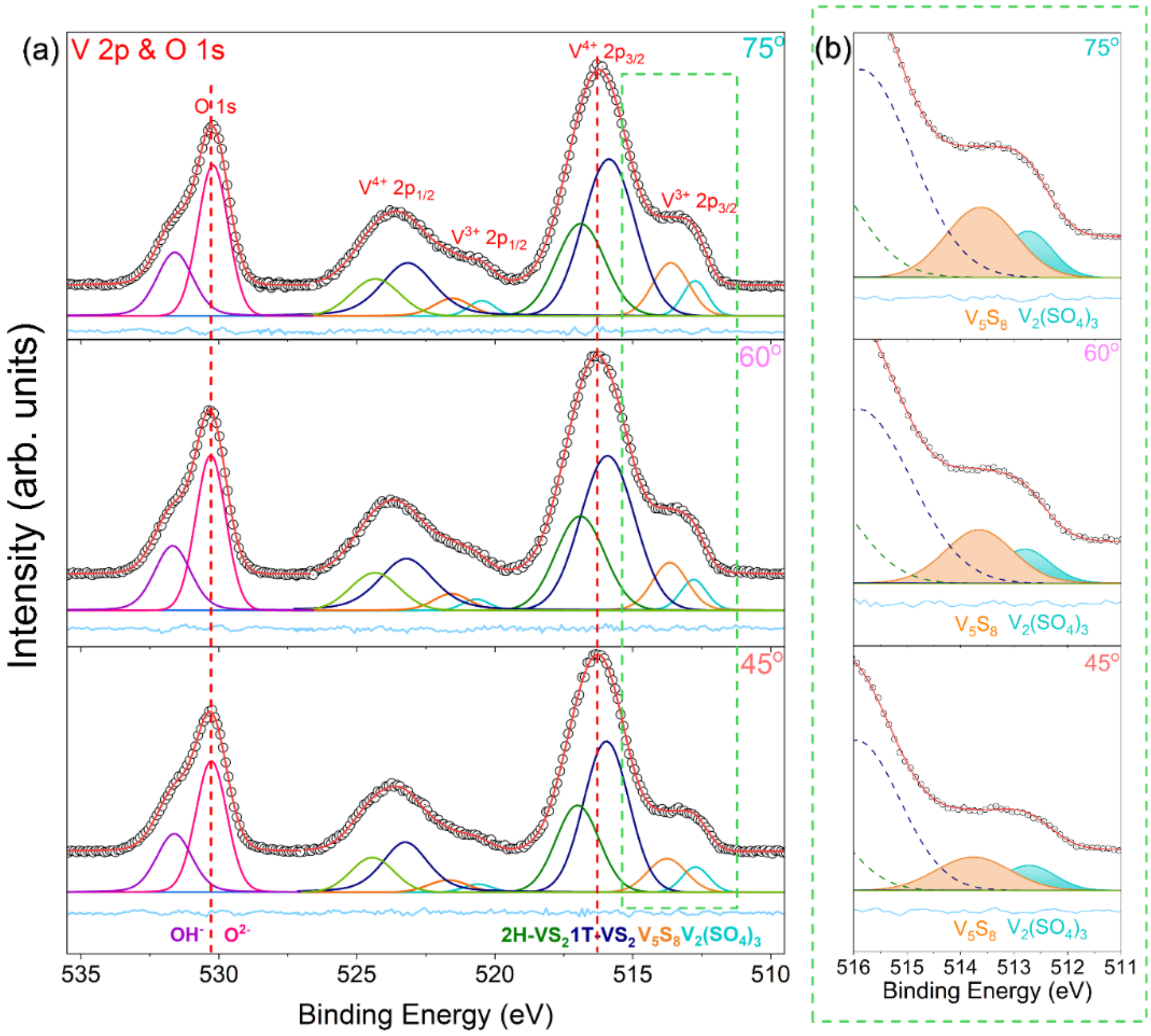

**FIGURE S7:** V 2p and O 1s core-level X-ray photoelectron spectra (black open circles) for the samples with different emission angles. Iterative Shirley background subtraction has been applied to all spectra. (a) Fitted spectra of V 2p and O 1s region (red solid line) taken at increasing emission angles, (b) Enlarged view of (a), highlighting the increase in area of $V_5S_8$ relative to $V_2(SO_4)_3$.

## 12. Fitting parameters of the V $2p_{3/2}$ spectra at different emission angles

| V $2p_{3/2}$ | | | | | | | | Acquisition Angle (degree) | Area Ratio $V^{3+}/V^{4+}$ | Area Ratio $V_5S_8$/ $V_2(SO_4)_3$ |
|---|---|---|---|---|---|---|---|---|---|---|
| $V^{3+}$ | | | | $V^{4+}$ | | | | | | |
| $V_2(SO_4)_3$ | | $V_5S_8$ | | 1T-$VS_2$ | | 2H-$VS_2$ | | | | |
| Peak Position (eV) | FWHM (eV) | Peak Position (eV) | FWHM (eV) | Peak Position (eV) | FWHM (eV) | Peak Position (eV) | FWHM (eV) | | | |
| 512.95 | 1.15 | 513.99 | 1.64 | 516.18 | 1.92 | 517.22 | 1.84 | 45 | 0.16 | 1.92 |
| 512.79 | 1.13 | 513.85 | 1.54 | 515.97 | 2.02 | 517.02 | 1.91 | 60 | 0.19 | 2.08 |
| 512.85 | 1.09 | 513.79 | 1.54 | 515.98 | 2.11 | 516.97 | 2.02 | 75 | 0.22 | 2.22 |

**Table-S5**: Fitting parameters of the V $2p_{3/2}$ spectra at different emission angles, highlighting the area ratios of $V^{3+}/V^{4+}$ and $V_5S_8$/ $V_2(SO_4)_3$.

## 13. Charge transfers in the unit (2 × 2) supercell of multilayer $VS_2$

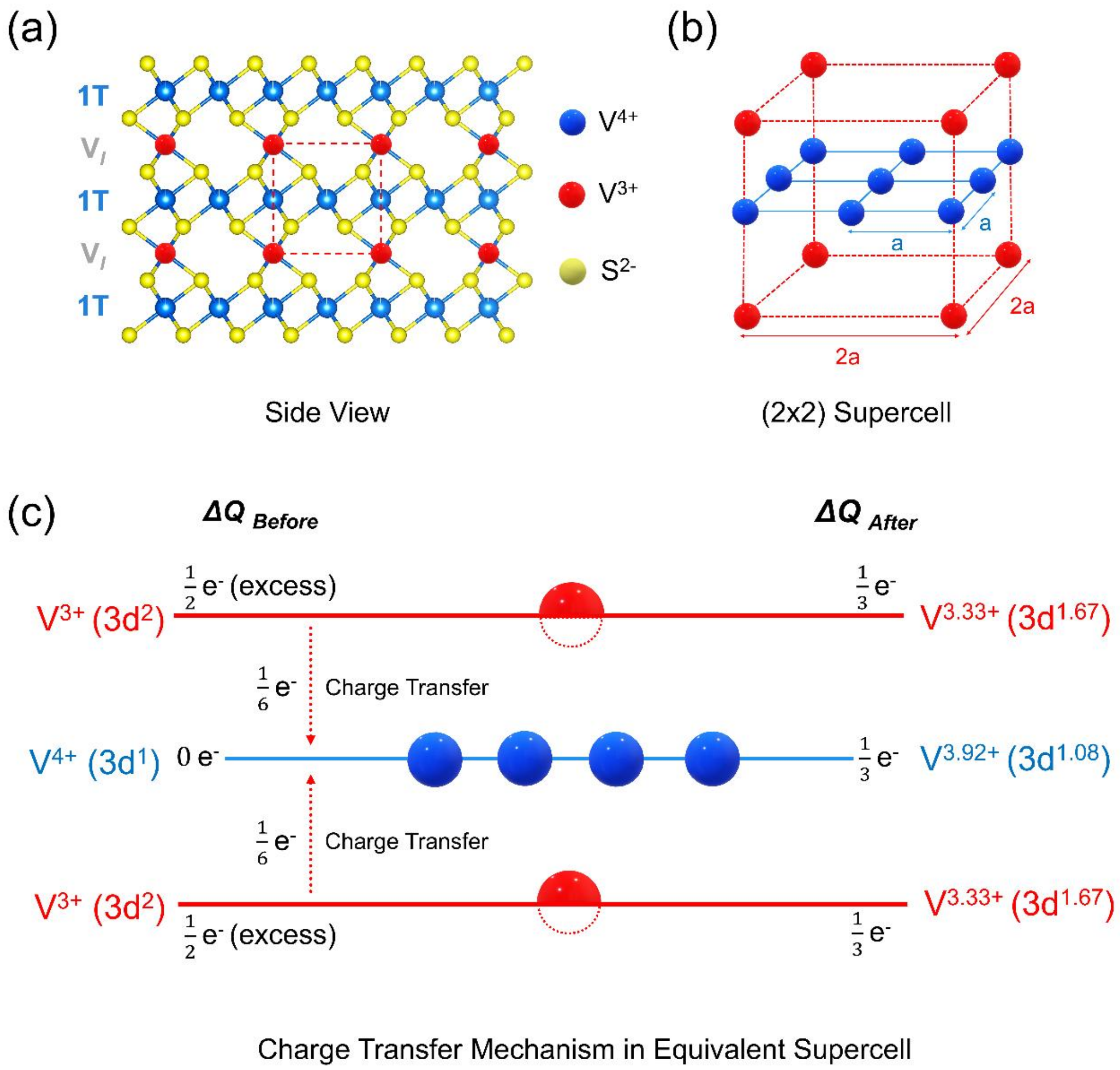

**FIGURE S8:** Charge transfers in the unit (2 × 2 × 1) supercell of multilayer $VS_2$. Figure S 8(a) shows a schematic of the $V^{3+}$ ion intercalations between the layers of a multilayer $VS_2$ forming the (2 × 2) superstructure in the basal plane. A unit supercell is marked by a red dotted square. Figure S 8(b) shows the schematic of a unit supercell in a 3D. The top and the bottom layers consist of $V^{3+}$ ions in (2 × 2) arrangements, and the middle layer consists of $V^{4+}$ ions arranged themselves with (1 × 1). The number of $V^{3+}$ ions in the unit supercell is one (half at the top layer and half at the bottom layer), and the $V^{4+}$ ions at the middle layer of the unit cell are 4. Due to this charge imbalance across the interlayer direction, charge transfer happens from the $V^{3+}$ sites to the $V^{4+}$ sites as there is one extra electron in the $V^{3+}$ ion than in the $V^{4+}$ ion. So, 1/2 electron is in excess in the top layer and bottom layer of each of the (2 × 2 × 1) unit cell. To balance the charge across the layers, 1/6 $e^-$ charge transfer happens from each $V^{3+}$ layer to the $V^{4+}$ layer, and the two lattices (1 × 1) and (2 × 2) charge balance to the one as shown in figure S 8(c).

## 14. XPS Survey and AFM phase contrast data

Figure S9 presents a standard X-ray photoelectron spectroscopy (XPS) survey spectrum of the synthesized $VS_2$, spanning a wide binding energy range. All fundamental levels and Auger characteristics have been determined to be associated with either vanadium or sulfur. This validates the sample's purity. The sample has been subjected to environmental conditions while being transported to the XPS experimental chamber. Consequently, a signal from ubiquitous contaminants like carbon and oxygen is also detectable.

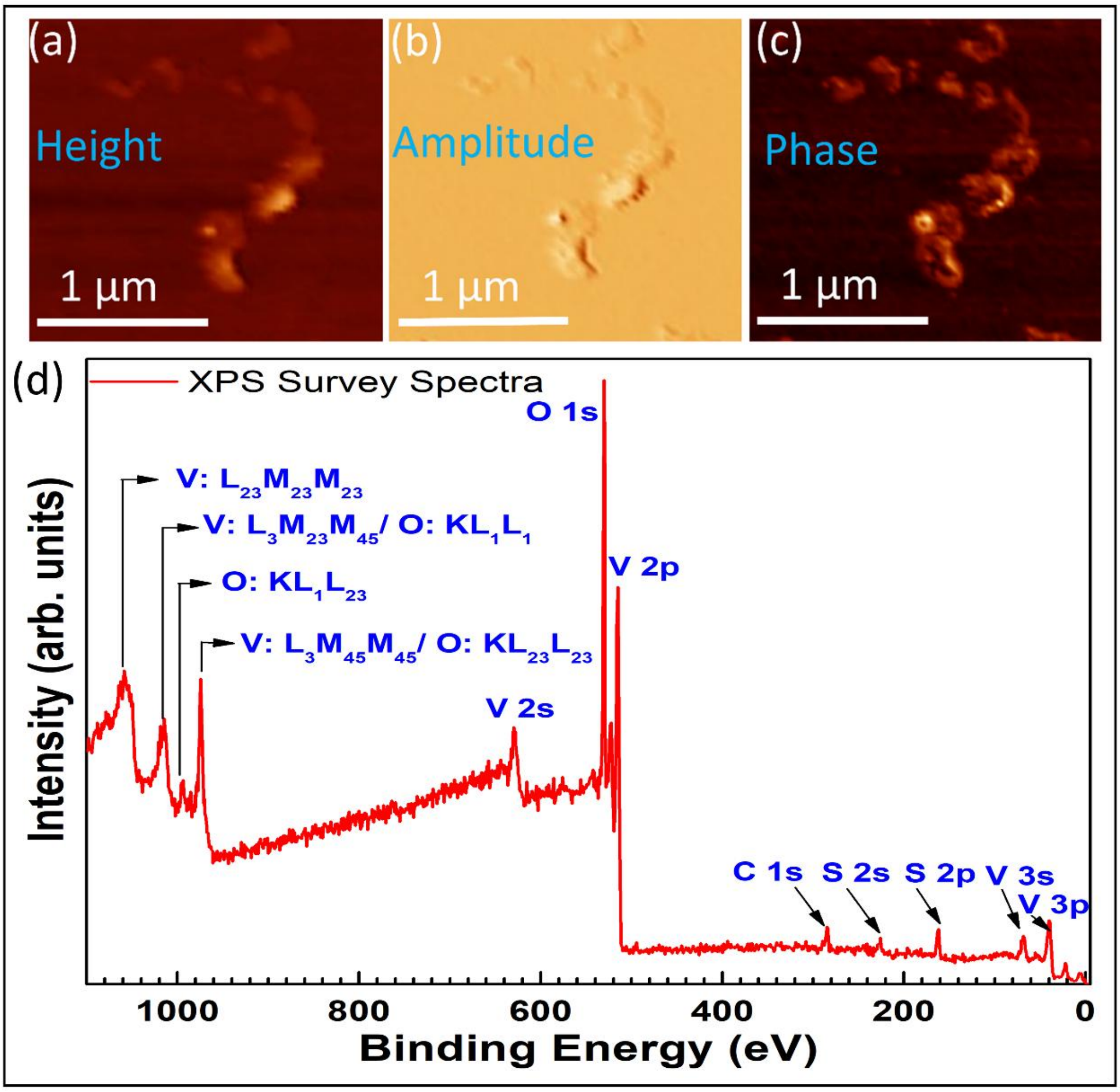

**Figure S9**: (a), (b), and (c) illustrate the AFM height contrast, amplitude, and phase contrast topographies of few-layer-thick $VS_2$ specimens acquired through the LPESTP technique. (d) displays a conventional XPS survey scan of the $VS_2$ specimen, displaying a broad binding energy spectrum. All features related to vanadium and sulfur have been identified and labelled.

## 15. Calibration of Vanadium oxidation states

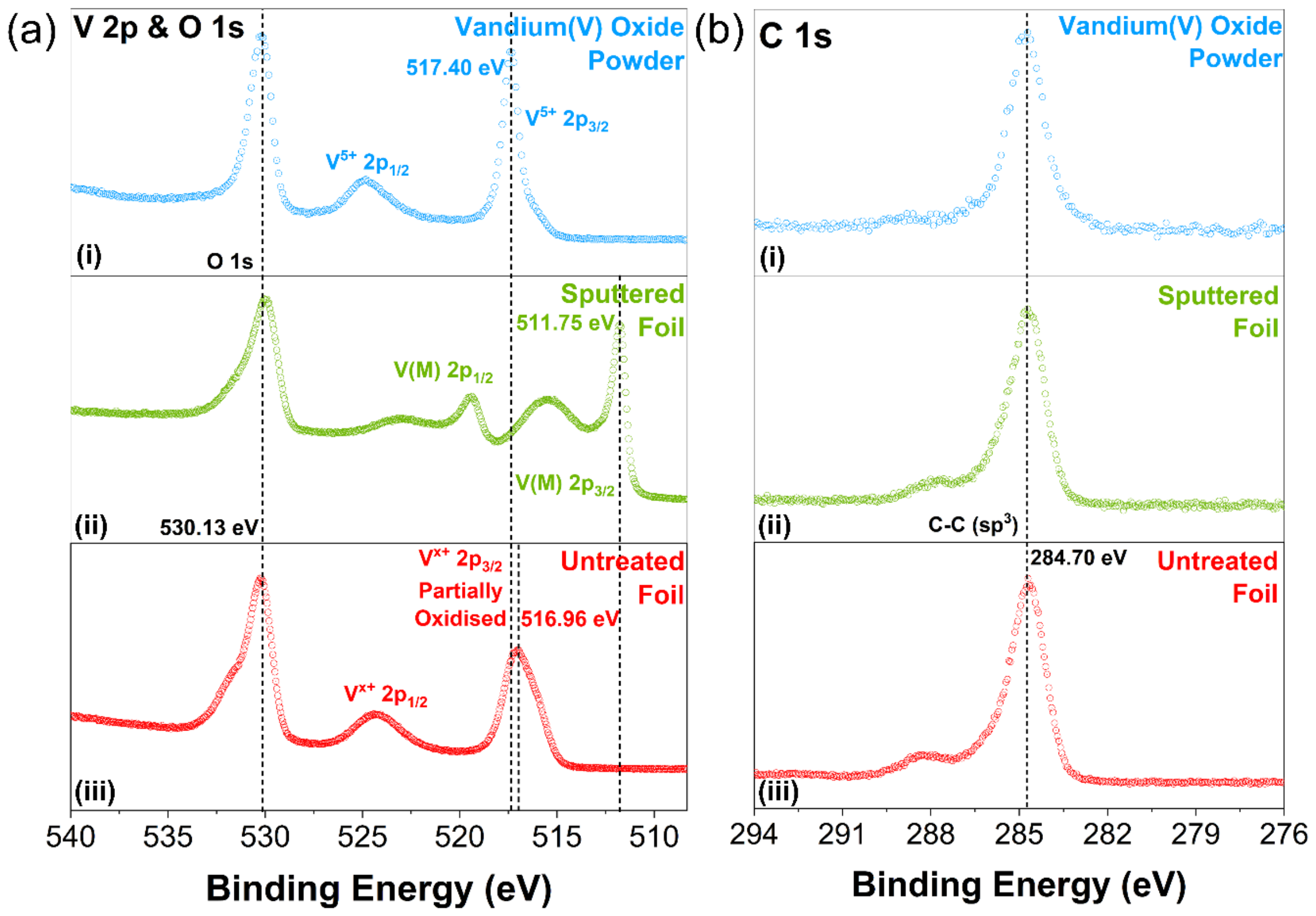

**Figure S10**: (a) V 2p and O 1s spectra for three samples are presented: (i) Vanadium (V) oxide ($V_2O_5$), which exhibits its highest oxidation state of +5. (ii) Vanadium metal foil analyzed after sputtering, indicating its metallic or zero oxidation state; and (iii) Vanadium metal foil examined before in situ $Ar^+$ ion sputtering, which displays an oxidation state represented as x ($+5 \geq x \geq +3$). All O 1s peaks are centered around 530.13 eV. (b) C 1s spectra for the three samples, presented in the same order, are charge-corrected to the C-C ($sp^3$) peak at 284.70 eV.

Before measurement, the vanadium foil was secured to the platen using perforated stainless steel sheets fastened with screws, effectively clamping the foil in place. The Vanadium (V) oxide powder was then compressed into pellets and attached to the sample using copper tabs. Surface decontamination was performed with an $Ar^+$ ion gun at a low accelerating voltage (300-500 V) for approximately 30 seconds. In situ sputtering of the metal foil was conducted at an accelerating voltage of 1 kV for a duration of 1 minute, with 30-second intervals. The spectra for all samples were measured at a consistent pass energy of 11.57 eV, utilizing an identical dual-beam charge-neutralization protocol. Additionally, the spectra from all three samples were charge-corrected to the C-C ($sp^3$) peak at 284.70 eV, as illustrated in Figure S 10(b). This correction enables a direct comparison of the spectral positions (V $2p_{3/2}$) across all three samples.

As shown in Figure S 10(a), metallic vanadium species have an asymmetrical peak profile and a narrow FWHM at 511.75 eV. This peak is lower by 1.2 and 2.24 eV compared to the peaks assigned to $V^{3+}$ of $VS_2$, which are located at 512.95 eV and 513.99 eV. Our analyzer's lower resolution limit is 300 meV (at 1486.7 eV photon energy), which is about one-third of the smallest difference value (1.2 eV), so the peaks at 512.95 and 513.99 eV are not metallic species. Detecting the $V^{5+}$ species at 517.40 eV shows that Vanadium's highest and lowest oxidation states are 5.65 eV apart. Our $VS_2$ peak assignments all fall within this range, supporting $V^{4+}$ and $V^{3+}$ split-charge-state assignments.

The parameters obtained from the peak-fitting of V 2p spectra taken from $VS_2$ are presented in Table S4. The spin-orbit splitting values for the $V^{3+}$ species are 7.88 and 7.92 eV, which are nearly identical. For the $V^{4+}$ species, the values are 7.29 and 7.44 eV; these are also quite similar and fall within the spectral resolution of the analyzer. Additionally, the full width at half maximum (FWHM) for the doublet pairs of the four assigned peaks are as follows: (1.15-1.19) eV, (1.64-1.67) eV, (1.92-1.95) eV, and (1.84-1.83) eV.

# 16. Temperature-dependent resistivity and magnetic transport data

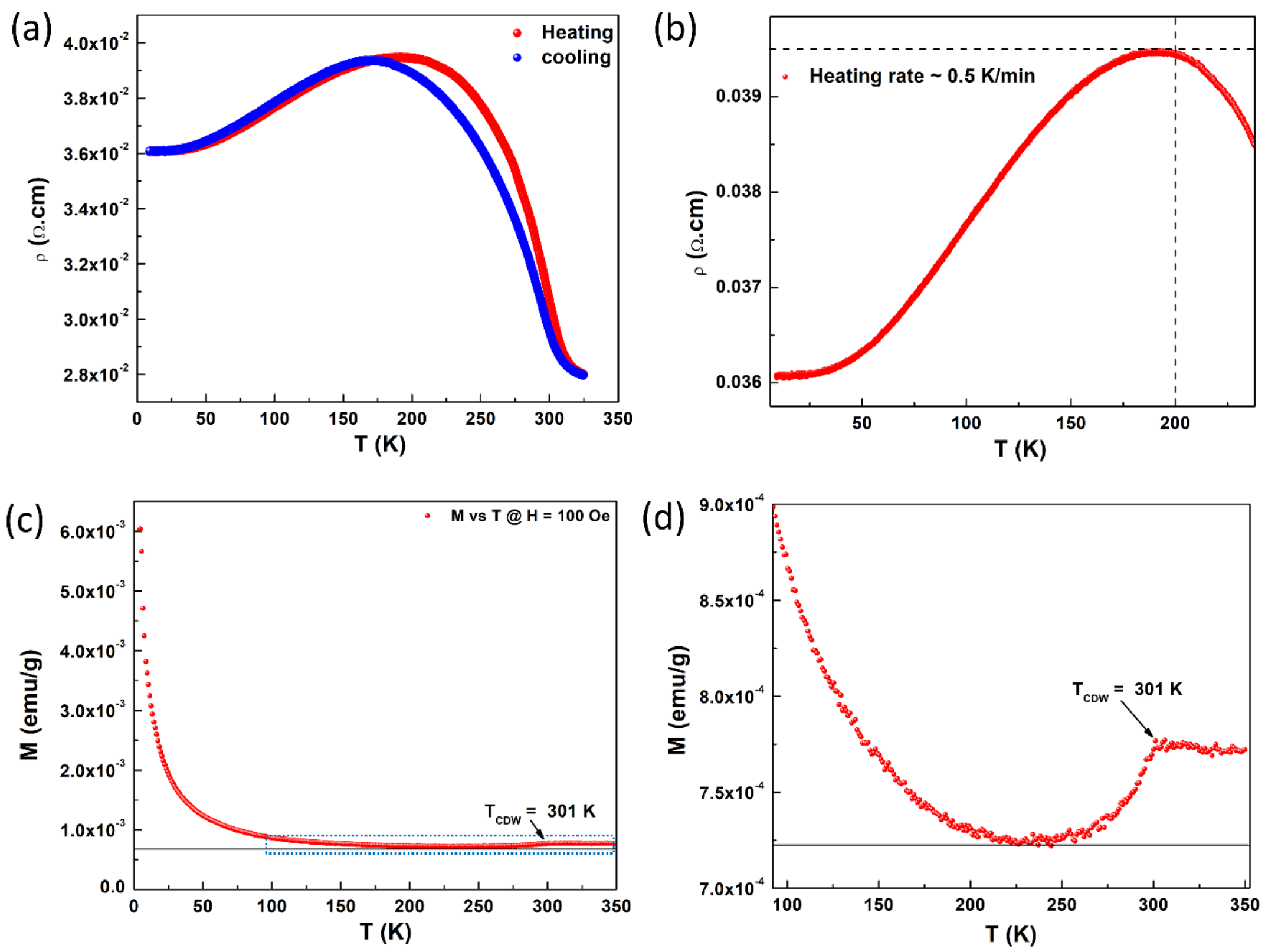

**Figure S11**: (a) shows the resistivity heating and cooling curve from 9 to 325 K at a ramp rate of 0.5 K/min. The resistance hysteresis observed between 175 and 325 K indicates a first-order CDW phase transition ($T_{CDW}$). (b) depicts the resistivity heating curve ranging from 9 to 245 K, emphasizing the metallic properties observed below 200 K. (c) displays the magnetization as a function of temperature (M-T) graphs for the multilayer $VS_2$ samples ranging from 2 to 350 K, highlighting the CDW transition temperature ($T_{CDW}$) at 301 K. (d) offers an enlarged perspective of the M−T curve from 90 to 350 K, as shown in (c), and illustrates that magnetization rises with power law when lowering the temperature below 225 K.

The temperature-dependent SAED measurements were conducted during in situ cooling and heating cycles (294 K → 218 K → 123 K → 294 K). The satellite reflections disappeared when the temperature dropped below 220 K and fully reappeared as the temperature returned to room temperature. This observation supports the conclusion of a reversible thermodynamic phase transition rather than sample deterioration. The transition temperature for the CDW in 1T-$VS_2$ is approximately 300 K, suggesting that the material should retain its CDW state below this temperature unless affected by competing phases. The temperature-dependent CDW phase transition is reversible. This is supported by our temperature-dependent selected area electron diffraction (SAED) patterns (Figure 2 and Figure S12), resistivity measurements, and magnetic transport investigations, as illustrated in Figure S11.

Moreover, it is known that $VS_2$ demonstrates ferromagnetism at room temperature and antiferromagnetism at cryogenic temperatures [14, 15, 16]. As the temperature decreases, magnetic properties become more pronounced, while the CDW phases tend to diminish. Consequently, the disappearance of the CDW superlattice spots when cooling below the established transition temperature is not unusual in $VS_2$ materials, which exhibit magnetic and other correlated phenomena at lower temperatures. The free-energy landscape of 2D-TMDCs with partially occupied *d*-bands has competing quantum ground states [17, 18]. Lower temperatures in monolayer 1T-$VS_2$ increase short-range electronic correlations and localized magnetic moments, which can be ferromagnetic or antiferromagnetic [19, 20]. Magnetic or strongly correlated states change electronic screening and reduce in-plane periodic lattice distortion, causing CDW satellites to melt at lower temperatures.

## 17. Convergence Benchmarks and Supercell Stress Criteria:

We have performed clear convergence evaluations for the large supercells. The production calculations employ a 600 eV plane-wave cut-off and an electronic convergence criterion of $10^{-8}$ eV, ensuring uniform reciprocal-space *k*-point densities throughout different supercells. The finite-displacement phonon calculations utilize a (4 × 4 × 1) (48-atom) supercell for pure 1T-$VS_2$ and a (2 × 2 × 1) (84-atom) supercell for the ($\sqrt{7} \times \sqrt{7}$) R19.1° arrangements, applying a displacement amplitude of 0.01 Å. The convergence table displayed below monitors the comparative CDW energies, soft-mode frequencies, peak residual forces, and in-plane residual stresses.

**Table-S6**. Numerical convergence of the competing CDW states

| Test | Parameter | ΔE(4×4) (meV/f.u.) | ΔE(√7×√7) (meV/f.u.) | ω(Q1) ($cm^{-1}$) | ω(Q2) ($cm^{-1}$) | Fmax (eV/Å) | \|σ\|\|max (kbar) |
|---|---|---|---|---|---|---|---|
| Cutoff | 500 eV | -10.82 | -12.04 | -47.8 | -29.6 | 0.012 | 0.46 |
| | 550 eV | -11.05 | -12.39 | -49.9 | -31.1 | 0.009 | 0.31 |
| | 600 eV (Our work) | -11.16 | -12.57 | -50.7 | -31.8 | 0.006 | 0.19 |
| | 650 eV | -11.16 | -12.57 | -50.7 | -31.8 | 0.006 | 0.19 |
| k-point density | coarse | -10.91 | -12.22 | -46.9 | -28.7 | 0.008 | 0.27 |
| | medium | -11.10 | -12.48 | -49.8 | -31.0 | 0.007 | 0.22 |
| | Our work | -11.16 | -12.57 | -50.7 | -31.8 | 0.006 | 0.19 |
| | denser | -11.16 | -12.57 | -50.7 | -31.8 | 0.006 | 0.19 |
| Force criterion | 0.020 eV/Å | -11.09 | -12.46 | -50.7 | -31.1 | 0.018 | 0.34 |
| | 0.010 eV/Å | -11.14 | -12.54 | -50.2 | -31.8 | 0.009 | 0.19 |
| | 0.005 eV/Å | -11.14 | -12.54 | -50.2 | -31.8 | 0.009 | 0.19 |

## 18. Crystallographic Basis and q-Vectors: (√7 × √7) R19.1° CDW Superstructure

The 19.1° rotation denotes the geometric alignment of the superlattice basis vectors ($\mathbf{A}_1 = 2\mathbf{a}_1 + \mathbf{a}_2$, $\mathbf{A}_2 = -\mathbf{a}_1 + 3\mathbf{a}_2$) in relation to the original 1T-$VS_2$ basis vectors ($\mathbf{a}_1$ and $\mathbf{a}_2$), essential for the periodic tilting of 7-atom hexagonal clusters across the two-dimensional plane:

$\cos\theta = (\mathbf{A}_1 \cdot \mathbf{a}_1) / (|\mathbf{A}_1||\mathbf{a}_1|) = (2 + \cos 60°) / \sqrt{7} = 2.5 / \sqrt{7} \approx 0.9449$, thus $\theta \approx 19.1°$

Nonetheless, the minute interatomic forces responsible for cluster formation operate radially along the direct nearest-neighbour V–V bonding axis (**a**1, **a**2 (first nearest), **a**1+**a**2 (second nearest)), which align perfectly with the high-symmetry Γ–K directions in reciprocal space.

### Real-Space and Reciprocal Basis Vectors of the Pristine (1 × 1) Unit Cell of 1T-$VS_2$

For the hexagonal lattice of pristine 1T-VS2, the in-plane primitive real-space lattice vectors can be represented as follows:

$\mathbf{a}_1 = a\,(1,0)$,

$\mathbf{a}_2 = a\,(1/2, \sqrt{3}/2)$

The reciprocal lattice basis vectors $b_1$, $b_2$ satisfying $a_i \cdot b_j = 2\pi\delta_{ij}$ have an angle of 120° between them:

$\mathbf{b}_1 = \frac{4\pi}{\sqrt{3}a}(\sqrt{3}/2, -1/2)$,

$\mathbf{b_2} = \frac{4\pi}{\sqrt{3}a}(0,1)$

## Real-Space Supercell Basis Vectors

The length condition for the supercell vector **A**= m $\mathbf{a_1}$ + n $\mathbf{a_2}$ with a 60◦ basis is:

$|A|^2 = a^2(m^2 + m.n + n^2) = 7a^2$

The two integer pairs producing a rotation of ±19.1◦ are (m,n) = (2,1) and (m,n) = (1,2):

Clockwise domain (+19.1◦):

$$\begin{pmatrix} \boldsymbol{A1} \\ \boldsymbol{A2} \end{pmatrix} = \begin{pmatrix} 2 & 1 \\ -1 & 3 \end{pmatrix} \begin{pmatrix} \boldsymbol{a1} \\ \boldsymbol{a2} \end{pmatrix} \Rightarrow M = \begin{pmatrix} 2 & 1 \\ -1 & 3 \end{pmatrix}, \det(M) = 7$$

M is the supercell transformation matrix.

Counter-clockwise domain (−19.1∘):

$$\begin{pmatrix} \boldsymbol{A'_1} \\ \boldsymbol{A'_2} \end{pmatrix} = \begin{pmatrix} 2 & 1 \\ -1 & 3 \end{pmatrix} \begin{pmatrix} \boldsymbol{a1} \\ \boldsymbol{a2} \end{pmatrix} \Rightarrow M = \begin{pmatrix} 2 & 1 \\ -1 & 3 \end{pmatrix}, \det(M) = 7$$

*(Both $A_1$ and $A_2$ maintain $|\mathbf{A_1}| = |\mathbf{A_2}| = 7a$ with a 60◦ angle between them: $\mathbf{A_1} \cdot \mathbf{A_2} = 27a^2$.)*

## Reciprocal Space Superstructure q-Vectors

The supercell reciprocal vectors $\mathbf{B}_i$ (which define the primary CDW modulation vectors) are obtained via the inverse transpose:

$$\begin{pmatrix} \boldsymbol{B1} \\ \boldsymbol{B2} \end{pmatrix} = (M^{-1})^T \begin{pmatrix} \boldsymbol{b1} \\ \boldsymbol{b2} \end{pmatrix}$$

1. For the +19.1° domain:

$$M^{-1} = \frac{1}{7}\begin{pmatrix} 3 & -1 \\ 1 & 2 \end{pmatrix} \Rightarrow (M^{-1})^T = \frac{1}{7}\begin{pmatrix} 3 & 1 \\ -1 & 2 \end{pmatrix}$$

This gives:

$\mathbf{B_1} = \frac{3}{7}\mathbf{b_1} + \frac{1}{7}\mathbf{b_2}$

$\mathbf{B_2} = -\frac{1}{7}\mathbf{b_1} + \frac{2}{7}\mathbf{b_2}$

The third symmetric in-plane wavevector forming the triple-q CDW state is:

$\mathbf{B_3} = -(\mathbf{B_1} + \mathbf{B_2}) = -\frac{2}{7}\mathbf{b_1} - \frac{3}{7}\mathbf{b_2}$

2. For the −19.1° domain:

$(M'^{-1})^T = \frac{1}{7}\begin{pmatrix} 2 & -1 \\ 1 & 3 \end{pmatrix}$

This gives:

$\mathbf{B_1}' = \frac{2}{7}\,\mathbf{b_1} - \frac{1}{7}\,\mathbf{b_2}$

$\mathbf{B_2}' = \frac{1}{7}\,\mathbf{b_1} + \frac{3}{7}\,\mathbf{b_2}$

$\mathbf{B_3}' = -(\mathbf{B_1}' + \mathbf{B_2}') = -\frac{3}{7}\,\mathbf{b_1} - \frac{2}{7}\,\mathbf{b_2}$

Magnitude: $|\mathbf{B}_i| = \frac{1}{\sqrt{7}}\,|\mathbf{b}_1| = \frac{4\pi}{\sqrt{21}\,a} \approx 0.378\,|\mathbf{b}|$

Angle between $\mathbf{B_1}$ and $\mathbf{B_2}$: 120°

Each structural domain above yields 6 modulation peaks (± $\mathbf{B_1}$, ± $\mathbf{B_2}$, ± $\mathbf{B_3}$) lying at a radius of $\frac{1}{\sqrt{7}}|\mathbf{b}|$ from the Γ point, rotated by ±19.1° away from the pristine Bragg peak directions.

## Primary q-Vectors and High-Symmetry Projections

The primary q-vector in terms of the pristine reciprocal basis vectors is given by

$\mathbf{q}_1 = \frac{3}{7}\,\mathbf{b_1} + \frac{1}{7}\,\mathbf{b_2} = h\,\mathbf{b_1} + k\,\mathbf{b_2}$

Where h and k are scalar fractions, $h = \frac{3}{7} = 0.428$ and $k = \frac{1}{7} = 0.143$.

Or the q-vector in terms of the scalar fractions is given by,

$q_1 = (h, k) = (\frac{3}{7}, \frac{1}{7}) = (0.428, 0.143)$

Chiral orientation of the modulation vector $\mathbf{q_1}$: 19.1° off nearest Γ–M and 10.9 ° off nearest Γ-K

Projections (parallel) along Γ-M ( $|\mathbf{\Gamma\text{-}M}| = \frac{b}{2} = \frac{2\pi}{a\sqrt{3}}$ ): $q_1 \cos(19.1^{\circ}) \approx (\frac{5}{14})\,b = \frac{5}{7}\,|\mathbf{\Gamma\text{-}M}|$

Therefore, the fraction along the path ≈ 0.71

Projection (parallel) along Γ-K ( $|\mathbf{\Gamma\text{-}K}| = \frac{b}{\sqrt{3}} = \frac{4\pi}{3a}$ ) : $q_1 \cos(10.9^{\circ}) \approx (\frac{3\sqrt{3}}{14})\,b = \frac{9}{14}\,|\mathbf{\Gamma\text{-}K}|$

Consequently, the ratio along the path is ≈ 0.65

## 19. Temperature-dependent SAED Patterns:

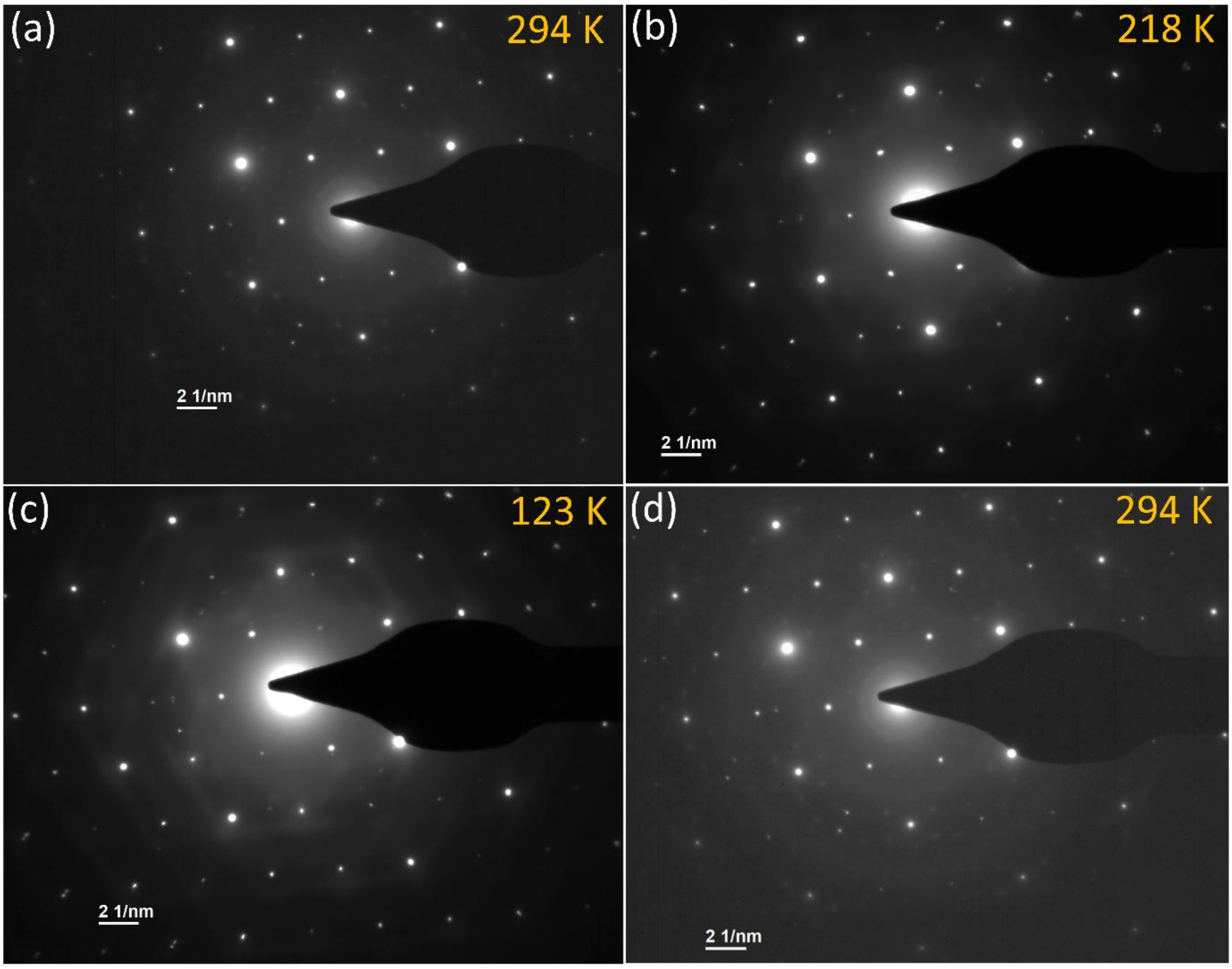

Figure S12: Temperature-dependent SAED patterns to observe the evolution of the monolayer (√7 × √7) R19.1° CDW superstructure. (a) SAED pattern at 294 K showing low-intensity (√7 × √7) R19.1° CDW superstructure spots along with bright pristine 1T-$VS_2$ Bragg Spots. (b) and (c) show the SAED patterns of the same flake at 218 K and 123 K, respectively, during cooling, which show no such (√7 × √7) R19.1° CDW superstructure spots at these temperatures and signify the complete melting of the CDW superstructure at these low temperatures. (d) shows the SAED pattern from the same flake after the temperature was raised to 294 K (room temperature) after reheating. The CDW super-spots, along with the bright pristine Bragg spots, clearly signify that the material returned to its (√7 × √7) R19.1° CDW phase near room temperature.